\documentclass[sigconf,anonymous=false]{acmart}
\usepackage[]{subfigure}
\usepackage{epstopdf}
\usepackage{booktabs} % For formal tables
\usepackage{textcomp}
\usepackage{booktabs}
\usepackage{tabularx}
\usepackage{courier}
\usepackage{listings}
\usepackage{xcolor}
\usepackage{color,soul}
\usepackage{multirow}
\usepackage{comment}
\usepackage{enumitem}
\usepackage{hyperref}
\usepackage[textsize=tiny]{todonotes}
\usepackage{algorithm}
\usepackage{algpseudocode}% http://ctan.org/pkg/algorithmicx
\algrenewcommand\textproc{}% Used to be \textsc
\usepackage[all]{nowidow}
\usepackage{array}
\usepackage{pifont}% http://ctan.org/pkg/pifont
\usepackage[skip=0pt,font=small]{caption}
\usepackage{bm}
\usepackage{tikz}
\usepackage{soul}
\usepackage{colortbl}

\providecommand{\DIFdel}[1]{} %Don't show deleted text
\usepackage{xcolor}
\definecolor{red}{RGB}{250,100,100}

\newcommand{\wahib}[1]{\textcolor{black}{#1}}
\newcommand{\leo}[1]{\textcolor{black}{#1}}
\newcommand{\nguyen}[1]{\textcolor{black}{#1}}
\newcommand{\albert}[1]{\textcolor{black}{#1}}

\usepackage[switch]{lineno} 

\newcommand{\rC}[0]{\rowcolor[HTML]{CCCCCC}}
\newcommand{\WC}[0]{\cellcolor[HTML]{CCCCCC}}

\newcommand\sbullet[1][.5]{\mathbin{\vcenter{\hbox{\scalebox{#1}{$\bullet$}}}}}
\newcommand\scirc[1][.5]{\mathbin{\vcenter{\hbox{\scalebox{#1}{$\circ$}}}}}
\newcolumntype{g}{>{\columncolor{lightgray}}l}
\newcolumntype{n}{>{\columncolor{lightgray}}c}

\newcommand{\aveacu}{86.74\% }
\newcommand{\maxacu}{97.57\% }

\newcommand{\bestpar}{data parallelism}

\copyrightyear{2021}
\acmYear{2021}
\setcopyright{acmcopyright}\acmConference[HPDC '21]{Proceedings of the 30th International Symposium on High-Performance Parallel and Distributed Computing}{June 21--25, 2021}{Virtual Event, Sweden}
\acmBooktitle{Proceedings of the 30th International Symposium on High-Performance Parallel and Distributed Computing (HPDC '21), June 21--25, 2021, Virtual Event, Sweden}
\acmPrice{15.00}
\acmDOI{10.1145/3431379.3460644}
\acmISBN{978-1-4503-8217-5/21/06}


\begin{document}
%\linenumbers

\title{An Oracle for Guiding Large-Scale Model/Hybrid Parallel Training of Convolutional Neural Networks}
\author{Albert Njoroge Kahira}
\authornote{Both authors contributed equally to this paper.}
\affiliation{
	\institution{Barcelona Supercomputing Center}
	\institution{ Universitat Politècnica de Catalunya}
	\state{Barcelona}
	\country{Spain}
}
\email{albert.kahira@bsc.es}

\author{Truong Thao Nguyen}
\authornotemark[1]
\affiliation{
	\institution{National Institute of Advanced Industrial Science and Technology}
	\country{Japan}
}
\email{nguyen.truong@aist.go.jp}

\author{Leonardo Bautista Gomez}
\affiliation{
	\institution{Barcelona Supercomputing Center}
	\state{Barcelona}
	\country{Spain}
}
\email{leonardo.bautista@bsc.es}

\author{Ryousei Takano}
\affiliation{
	\institution{National Institute of Advanced Industrial Science and Technology}
	\country{Japan}
}
\email{takano-ryousei@aist.go.jp}

\author{Rosa M Badia}
\affiliation{
	\institution{Barcelona Supercomputing Center}
	\institution{ Universitat Politècnica de Catalunya}
	\state{Barcelona}
	\country{Spain}
}
\email{rosa.m.badia@bsc.es}

\author{Mohamed Wahib}
\affiliation{
	\institution{National Institute of Advanced Industrial Science and Technology}
	\institution{RIKEN-CCS}
	\country{Japan}
}
\email{mohamed.attia@aist.go.jp}

\renewcommand{\shortauthors}{Albert Kahira, Truong Thao Nguyen et al.}

\begin{abstract}
	Deep Neural Network (DNN) frameworks use distributed training to enable faster time to convergence and alleviate memory capacity limitations when training large models and/or using high dimension inputs. With the steady increase in datasets and model sizes, model/hybrid parallelism is deemed to have an important role in the future of distributed training of DNNs. We analyze the compute, communication, and memory requirements of Convolutional Neural Networks (CNNs) to understand the trade-offs between different parallelism approaches on performance and scalability. We leverage our model-driven analysis to be the basis for an oracle utility which can help in detecting the limitations and bottlenecks of different parallelism approaches at scale. We evaluate the oracle on six parallelization strategies, with four CNN models and multiple datasets (2D and 3D), on up to 1024 GPUs. The results demonstrate that the oracle has an average accuracy of about \aveacu when compared to empirical results, and as high as \maxacu for \bestpar.
\end{abstract}

\maketitle

\section{Introduction} 
\label{sec:Introduction}

DNNs are achieving outstanding results in a wide range of applications, including image recognition, video analysis, natural language processing~\cite{shen2014learning}, understanding climate~\cite{hurricanes}, and drug discovery~\cite{wallach2015atomnet}, among many others.
In the quest to increase solution accuracy, researchers are increasingly using larger training datasets as well as larger and deeper DNN models~\cite{ben2018demystifying,you2018imagenet, huang2018gpipe}.
In addition, applying Deep Learning (DL) in new domains, such as health care and scientific simulations, introduce larger data samples and more complex DNN models~\cite{Kurth2018EDL}.
Those trends make the DNN training computationally expensive for a single node. Therefore, large-scale parallel training on high-performance computing (HPC) systems or clusters of GPUs is becoming increasingly common to achieve faster training times for larger models and datasets~\cite{ben2018demystifying}.
When training a specific DNN model on an HPC system, there are two prominent strategies for parallelizing the training phase of DL: \emph{data} and \emph{model} parallelism. It is important to note that despite the early investigation of model parallelism in DL~\cite{Dean2012DistributedDL}, those efforts were premature and remained far from production deployments, since data parallelism was simple and sufficient. However, the growth in datasets and models far outgrows the increase in compute capability~\cite{huang2018gpipe}. Accordingly, scaling data parallelism can be limited by the memory capacity and the communication overhead. 
\textit{First,} we elaborate on the memory capacity issue. In data parallelism, the entire model is duplicated for each compute node. Therefore, training larger and deeper neural networks have to deal with the memory capacity limits.
A notable case is in the area of language modeling at which models are increasingly approaching $O(100B)$ parameters~\cite{Rajbhandari2019ZeROMO} (ex: GPT-3 has 175B parameters~\cite{brown2020language}). In addition, for sample sizes with higher dimensions, e.g., 3D scientific data sets~\cite{CosmoflowDataset} and videos, the memory capacity would also limit the number of samples that can be concurrently processed by a GPU~\cite{Mathuriya:2018:CUD:3291656.3291743} hence, restricting the scaling of data parallelism. \textit{Second,} we elaborate on the communication overhead. A major bottleneck when scaling data parallelism is the large-message Allreduce collective communication for the gradient exchange at the end of each iteration~\cite{jia2018highly, yamazaki2019yet}.
Several active efforts try to optimize the Allreduction collective algorithm for supporting large messages on specific network architectures in HPC systems~\cite{ yamazaki2019yet,bayatpour2017scalable,nguyen2018hierarchical}. 
However, even with those algorithms, communication remains a bottleneck when the size of the models increases. Moving to model or hybrid parallelism is one of the ways to reduce this communication overhead~\cite{Naoya:sc19}.

\begin{table*}[bt]
	\caption{\small{Recent progress (and examples) in scaling distributed training of DNNs, across different training components and phases. Training components: \textbf{AP}- application (models and datasets), \textbf{TA}-training algorithms, \textbf{PS}-parallel strategies (computation and communication), \textbf{FR}-framework and \textbf{SY}-computer systems.
	Training phases: \textbf{IO}-I/O and pre-processing, \textbf{FB}-a forward and backward propagation, \textbf{GE}-the gradient exchange (if needed) and \textbf{WU}-updating the weights.
	$\sbullet[1.5]$: related components. $\checkmark$: related training phases.
	Explanation in remarks.}}
	\label{table:relatedwork}
	\centering
	\setlength\tabcolsep{2pt} % default value: 6pt
	\resizebox{\textwidth}{!}{%
	\begin{tabular}{lcccccccccl}
		\toprule
		\multirow{2}{*}{\textbf{\begin{tabular}[c]{@{}c@{}}Approaches\end{tabular}}} &
		\multicolumn{5}{c}{\textbf{\begin{tabular}[c]{@{}c@{}}Components\end{tabular}}} &
		\multicolumn{4}{c}{\textbf{Training Phases}} &
		\multirow{2}{*}{\textbf{Additional Remarks}} \\ \cmidrule(lr){2-6} \cmidrule(lr){7-10}
		& \textbf{AP} & \textbf{TA} & \textbf{PS}  & \textbf{FR} & \textbf{SY}  &\textbf{IO} & \textbf{FB} & \textbf{GE} & \textbf{WU} &  \\ \midrule
		\begin{tabular}[c]{@{}l@{}} Optimization methods
		\end{tabular} & $\scirc[1.5]$ & $\sbullet[1.5]$  & $\scirc[1.5]$ & $\scirc[1.5]$ & $\scirc[1.5]$ &  -  &  - &  - &   $\checkmark$ &  
		\begin{tabular}[c]{@{}l@{}} Second order methods~\cite{pauloski2020convolutional} (fewer epochs to converge, but longer iterations and more memory).
		\end{tabular}\\ \rC
		Normalization & $\sbullet[1.5]$ & $\scirc[1.5]$  & $\scirc[1.5]$ & $\scirc[1.5]$ & $\scirc[1.5]$ &  -  &  $\checkmark$ &  - &   - &
		\begin{tabular}[c]{@{}l@{}} Cross-GPU Batch-normalization~\cite{Zhang_2018_CVPR} (requires extra communication) and Group Normalization~\cite{wu2018group}.
		\end{tabular}\\
		Pre-trained model  & $\sbullet[1.5]$ & $\sbullet[1.5]$  & $\scirc[1.5]$ & $\scirc[1.5]$ & $\scirc[1.5]$ &  -  &  $\checkmark$ &  - &  - &
		\begin{tabular}[c]{@{}l@{}} A big model, that is pre-trained on a big generic dataset, such as Google BiT~\cite{alex2019big} can be fine-tuned for any task, even if only few \\labeled samples are available.
		\end{tabular}\\ \rC
		\begin{tabular}[c]{@{}l@{}} Allreduce optimization
		\end{tabular} & $\scirc[1.5]$ & $\scirc[1.5]$  & $\sbullet[1.5]$ & $\scirc[1.5]$ & $\sbullet[1.5]$ &  -  &  - &  $\checkmark$ &  - &
		\begin{tabular}[c]{@{}l@{}} Reduce the communication time by considering the specific network architectures of HPC systems~\cite{ yamazaki2019yet,bayatpour2017scalable,nguyen2018hierarchical} (data parallelism)
		\end{tabular}\\	
		Sparsification & $\scirc[1.5]$ & $\sbullet[1.5]$  & $\sbullet[1.5]$ & $\scirc[1.5]$ & $\scirc[1.5]$ & - &  $\checkmark$ &  $\checkmark$ & $\checkmark$ &
		\begin{tabular}[c]{@{}l@{}} Reducing the computation volume~\cite{Lym_PruneTrain} or/and communication message size~\cite{renggli2018sparcml} by skipping the computing/transferring of non-\\important weights/gradients, instead only performing on the significant ones (mainly for data parallelism).
		\end{tabular}\\ \rC
		Memory optimization & $\scirc[1.5]$ & $\scirc[1.5]$  & $\scirc[1.5]$ & $\sbullet[1.5]$ & $\scirc[1.5]$ &  -  &  $\checkmark$ &  $\checkmark$ &  $\checkmark$ &
		\begin{tabular}[c]{@{}l@{}} Reduce required memory by using lower precision (quantization)~\cite{1-bit-SGD}, gradient checking point \cite{Chen2016TrainingDN}, out-of-core methods~\cite{wahib2020scaling},

		\end{tabular}\\
		Network architecture & $\scirc[1.5]$ & $\scirc[1.5]$  & $\scirc[1.5]$ & $\scirc[1.5]$ & $\sbullet[1.5]$ &  $\checkmark$  &  - &  $\checkmark$ &   - &
		\begin{tabular}[c]{@{}l@{}}Increasing number of GPUs intra node, i.e., up to 16 GPUs in DGX-2. %Large-bandwidth divides such as NVLINK, NVSwitch to \\connect GPUs, CPU (in Summit) together. 
		High-throughput inter-node network topology such as \\HyperX~\cite{Domke_HyperX} or BiGraph~\cite{EFLOPS}. %In-network computing performs the reduction operation in switches (cite intel? ). 
		\end{tabular}\\
		\midrule
		\begin{tabular}[c]{@{}l@{}}\textbf{Model/hybrid parallelism} \\ (our target in this work)
		\end{tabular}& $\scirc[1.5]$ & $\scirc[1.5]$  & $\sbullet[1.5]$ & $\scirc[1.5]$ & $\scirc[1.5]$ &   $\checkmark$  &   $\checkmark$ &   $\checkmark$ &   $\checkmark$ &
		\begin{tabular}[c]{@{}l@{}}	Castello et al.~\cite{Castello:2019:AMP:3343211.3343218} analyzed the communication trade-offs in some model parallel strategies. Hybrid parallelism are proposed in\\~\cite{gholami2017integrated} (spatial with data) and~\cite{Naoya:sc19} (channel/filter with data). \cite{jia2018exploring,jia2018beyond} explored different parallelization schemes on per-layer basis. 
		\end{tabular}\\	 
%				AAA & $\scirc[1.5]$ & $\scirc[1.5]$  & $\scirc[1.5]$ & $\scirc[1.5]$ & $\scirc[1.5]$ &  -  &  - &  - &   - &
%\begin{tabular}[c]{@{}l@{}}
%\end{tabular}\\	 
		\bottomrule
	\end{tabular}
	}
\vspace{-0.3cm}
\end{table*}

Given those issues with data parallelism and the growing scale of training, researchers are tackling different bottlenecks across the different components necessary for distributed DNN training. Table~\ref{table:relatedwork} shows a summary of the recent approaches in scaling distributed DNN training, split into components and training phases. \textbf{Despite those efforts, data parallelism is not feasible for all cases.} Thus, it is important to understand the limitations and scalability of large scale model and hybrid parallelism training of DNNs. 

In this work, we focus on the HPC aspects of scaling six different strategies for model and hybrid parallelism in CNNs distributed training. Innovations in DL theory (e.g., optimizers) are out of the scope of this paper. While most works in the literature focus on improving the performance of one single parallelism strategy for one specific framework; our study functions as the basis for a tool, named \emph{ParaDL}, capable of modeling and predicting the performance of a large set of configurations for CNN distributed training at scale. In addition, ParaDL also helps to reveal the practical limits and bottlenecks of different parallel strategies in CNN training .

Our main contributions in this work are as follow:
\begin{itemize}[leftmargin=3mm]
	\item We formally define the main parallel strategies  (See Section~\ref{sec:strategies}), including hybrid ones, and provide a comprehensive 
	analysis of the compute, communication, and memory footprint when training CNNs for inputs of any dimension. 
	\item We propose an oracle (ParaDL)~\footnote{\nguyen{\url{https://github.com/TruongThaoNguyen/paraDL-analysis}}} 
	that projects the ideal performance  
	of distributed training of CNNs, broken down by training phases. This helps in favoring a parallel strategy on a given system (See Section~\ref{sec:model}) and aids in identifying optimization opportunities in frameworks.
	\item We implement all parallelization strategies to validate our model, except for: a) data parallelism (already supported by most DL frameworks), and b) using an existing pipeline implementation.
	\item We show the utility of ParaDL in exposing performance and scalability trade-offs. The accuracy of ParaDL (\aveacu on average and up to \maxacu) is demonstrated by conducting a wide range of experiments for different CNN models, parallel strategies, and datasets, on up to $1,024$ GPUs (See Section~\ref{sec:evaluation}). 
\end{itemize}

\section{Background}
\label{sec:Background}	
\subsection{Phases of Distributed Training of DNNs}
\subsubsection{Stages of Distributed Training}
\label{sec:stages}
DNNs are made up of a network of neurons (represented as nodes) that are organized in layers (a model).
A DNN is trained by iteratively updating the weights of connections between layers in order to reduce the error in prediction of labelled datasets.
That is, for a given dataset of $D$ samples, a DNN is trained to find out the model weights $w$ for which the loss function $L$ is minimized.
Distributed training of a DNN can be divided into four phases: (\textbf{IO}) I/O and pre-processing, (\textbf{FB}) a forward phase at which the samples pass through the entire network, followed by a backward phase (back propagation) to compute the gradients, (\textbf{GE}) the gradient exchange (if needed) and (\textbf{WU}) updating the weights.
Specifically, the samples are picked up from the dataset randomly in batches of size $B$ (mini-batch).
The training process is then performed on those batches of samples iteratively by using an optimization algorithm such as the SGD, in which, weights are updated with a learning rate $\rho$ via $w^{iter+1} \gets w^{iter} - \rho\frac{1}{B} \sum_{i \in Batch}\Big(\frac{dL}{dw}\Big)_i$. The process is then repeated, until convergence, in epochs that randomize the order at which the input is fed to the network.

\subsubsection{Components of Distributed Training}
\label{sec:component}

To optimize for the performance and efficiency of training at large-scale, researchers introduce improvements to the methods, algorithms and design across entire training components which includes: \textbf{AP}- application (Deep Learning models and datasets), \textbf{TA}-training algorithms (ex: SGD or second order methods), \textbf{PA}-parallel strategies (model of computation and communication), \textbf{FR}-framework and \textbf{SY}-systems.

\subsection{Notation}
\label{sec:notation}
We summarize our notation in Table~\ref{table:notation}. In a $G$-layer CNN model, a convolution layer $l$ mainly needs these tensors:
\begin{itemize}[leftmargin=3mm]
	\item The input of layer $l$ with $N$ samples, each sample include $C_l$ channels, each channel is a tuple of $d$-dimension:  $x_l[N,C_l,X_l^{d}]$. In a 2-dimension layer, we replace $X_l^{d}$ with $[W_l, H_l]$, i.e., $x[N,C_l,W_l \times H_l]$. In a clear context, we omit the layer index $l$ and the dimension $d$, i.e., $x[N,C,X]$.
	\item The output (activation) of layer $l$ with $N$ samples and $F_l$ output channel $y_l[N,F_l,Y_l^{d}]$.
	\item The weight $w_l[C_l,F_l,K_l^{dim}]$ with $F_l$ filters. Each filter has $C_l$ channels and size of $K_l^{d}$. In some case, we omit the filter size (also known as kernel size), e.g., $w_l[C_l,F_l]$. 
	\item \nguyen{The bias $bi_l[F_l]$}.
	\item The activation gradients $\frac{dL}{dy_l}[N,F_l,Y_l^{d}]$.
	\item The weight gradients $\frac{dL}{dw_l}[C_l,F_l,K_l^{d}]$
	\item The input gradients $\frac{dL}{dx_l}[N,C_l,X_l^{d}]$.
\end{itemize}
%% Layer adaption
We adapt non-Conv layers with the above tensors (we extend the notation in~\cite{Naoya:sc19}). 
For channel-wise layers, such as pooling or batch-normalization, we require no further adaption.
A fully-connected layer with input $x[N,C,W \times H]$ and $F$ output can be expressed as a convolution layer where the size of filters is exactly similar to the size of the input layer, i.e., $w[C,F,W \times H]$, with padding and stride set to $0$ and $1$, respectively. Thus, the output will become $y[N,F,1 \times 1]$.
For element-wise layers, such as ReLU, the number of filters $F$ is equal to the number of channels $C$. 
For layers without weight, such as pooling and ReLU, the weight becomes $w[C,F,0]$.

The sequential implementation of CNN requires the following steps for each layer :
\begin{itemize}[]
	\small
	\item[(IO)] $x[N,C,X] \gets IO(\text{dataset, B})$ in the first layer %or\\
	%$x_l[N,C,X] \gets y_{l-1}[N,F,Y]$ if $l > 1$}
	\item[(FB)] $y[N,F,Y] \gets FW(x[N,C,X], w[C,F,K])$
	\item[(FB)] $\frac{dL}{dx}[N,C,X] \gets BW_{data}(\frac{dL}{dy}[N,F,Y], w[C,F,K])$
	\item[(FB)] $\frac{dL}{dw}[C,F,K] \gets BW_{weight}(\frac{dL}{dy}[N,F,Y], x[N,C,X])$
\end{itemize}
%The last layer calculates the loss on the output: 
%$\frac{dL}{dy}[N,F,Y] \gets L(y[N,F,Y])$.
At the end of each iteration, weights of all  layers are updated:
\begin{itemize}[]
	\small
	\item[(WU)]$w[C,F,K] \gets WU(\frac{dL}{dw}[C,F,K],\rho)$
\end{itemize}

\section{Strategies for Distributed Training}
\label{sec:strategies}
\begin{table}[]
	\centering
	\caption{Parameters and Notation}
	%\small 
	\scriptsize
	\begin{tabular}{|p{1.5cm}|p{6.2cm}|}
		\hline
		$D$ & Data set size \\ \hline
		$B$ & Mini batch size \\ \hline
		%$\rho$ & Learning rate\\ \hline
		$I$ & Number of iterations per epoch. $I=\frac{D}{B}$ \\ \hline
		$E$ & Number of epochs \\ \hline
		$G$ & Number of layers \\ \hline
		%lay$_i$ & $i^{th}$ layer in the neuron networks, $i \in 0 \dots L$ \\ \hline
		$x_l$ & Input of a layer $l$ \\ \hline
		$y_l$ & Output (activation) of layer $l$ \\ \hline
		$w_l$ & Weight of layer $l$ \\ \hline
		\nguyen{$bi_l$} & \nguyen{Biases of layer $l$} \\ \hline
		$W_l$ / $H_l$ & Width / Height of input of layer $l$ \\ \hline
		%& Height of input of layer $i$ \\ \hline
		$C_l$ & Number of input channels of layer $l$ \\ \hline
		$F_l$ & Number of output channels of layer $l$, e.g., number of filters in conv. layer\\ \hline
		%$T_{comp}$ & Total time for computation \\ \hline
		%$T_{comm}$ & Total time for communication \\ \hline
		$FW_l$ / $BW_l$ & Forward / Backward propagation action of layer $l$\\ \hline
		%$BW_l$ & Backward propagation action of layer $l$\\ \hline
		$[A_1,\dots,A_n]$ & $n$-dimensions array with size of $A_1 \times A_2 \times \dots \times A_n$\\ \hline  
		$X_l^{d}$ &  a $d$-dimension tuple (array) presents an input channel. In a 2-D convolution layer, $X_l^2$ is a Cartesian product of $W_l \times H_l$\\ \hline
		$Y_l^{d}$ &  \begin{tabular}[c]{@{}l@{}}a $d$-dimension output channel \end{tabular}\\ \hline
		$K_l^{d}$ & a $d$-dimension filter. In a 2-D convolution, $K_l^2 = K \times K$\\ \hline
		$p$ & Total number of processes elements (PEs)\\ \hline %$p=p_r \times p_c$\end{tabular} \\ \hline
		$S$ & Number of segments in pipeline parallelism \\ \hline
		$\alpha$ & \begin{tabular}[c]{@{}l@{}}Time for sending a message from source to destination \end{tabular} \\ \hline
		$\beta$ & Time for injecting one byte of message into network\\ \hline
		$\delta$ & Number of bytes per item, e.g., input, activation, weight\\ \hline
		$\gamma$ & Memory reuse factor\\ \hline
	\end{tabular}
	\label{table:notation}
	\vspace{-0.4cm}
\end{table}

Training CNNs using a single processing element (PE) is computationally expensive, e.g., training ResNet-50 over a single V100 GPU requires 29 hours. 
Hence distributed training on HPC systems is common for large models and datasets.
 Parallelizing of training process should be done by splitting different dimensions.
In this work, we cover four basic parallel strategies that differ in the way we split the data and model dimensions in the training of CNNs:
(1) distributing the data samples among PEs (\emph{data parallelism}), 
(2) splitting the data sample by its spatial dimension such as width or height (\emph{spatial parallelism})~\cite{ dryden2019improving},
(3) vertically partitioning the neural network along its depth (\emph{layer parallelism}) and overlapping computation between one layer and the next layer~\cite{huang2018gpipe} (also known as \emph{pipeline parallelism}), and
(4) horizontally dividing the neural network in each layer by the number of input and/or output channels (\emph{channel and filters parallelism})~\cite{gholami2017integrated, dryden2019improving}.
In addition, a combination of two (or more) types of the mentioned parallelism strategies is named as \emph{hybrid parallelism} (e.g., Data+Filter parallelism and Data+Spatial parallelism, or \textit{df} and \textit{ds} respectively for short, are some examples of hybrid parallelism).

In this work, when presenting tensors such as $x$, $y$, and $w$, we use the $*$ symbol to present a dimension for which its values are replicated between processes.
To emphasize that a tensor's dimension is partitioned among different PEs, we use the number of processes $p$.
For example, in data parallelism, $x[p,*,*]$ implies that the input $x$ is split equally in dimension $N$ (number of samples) and partitioned to $p$ PEs.
The other dimensions such as $C$ and $X$ are replicated.
The arrow $\xleftarrow[]{Allreduce}$ presents Allreduce communications. %operation.

It is important to note that the notation and analysis in this paper are general to input tensors of any dimension (1D, 2D, and, 3D). Input tensors of higher dimensions are also valid in our analysis since they can be represented as 3D tensor with the extra dimensions as component vector(s) (e.g. CosmoFlow~\cite{Mathuriya:2018:CUD:3291656.3291743} has 4D input represented as a 3D tensor plus a vector at each cell). Finally, the parallel strategy can alternatively be viewed as a domain decomposition problem: a recurring problem in HPC applications. Accordingly, we formulate the notation and analysis to be interpretable as domain decomposition schemes.
\begin{figure}[]
	\centering
	\small
	\subfigure [Sequential implementation on a single PE] {
		\label{fig:orig}
		\includegraphics[width=7.5cm,trim=1cm 0.2cm 1cm 0.2cm]{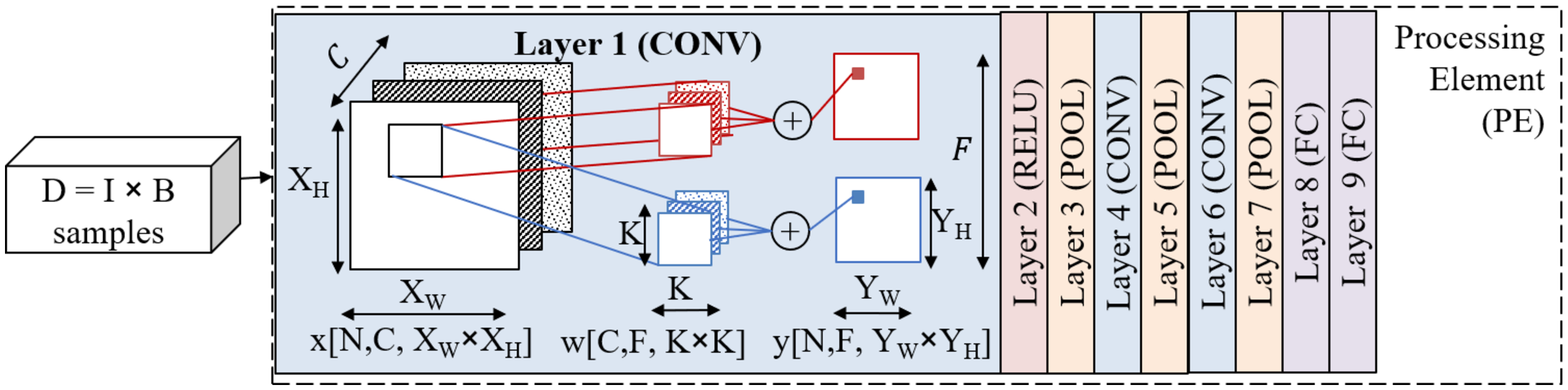}
	}
	\\
	\vspace{-0.2cm}
	\subfigure [Data parallelism] {
		\label{fig:data}
		\includegraphics[width=7.5cm,trim=1cm 0.2cm 1cm 0.2cm]{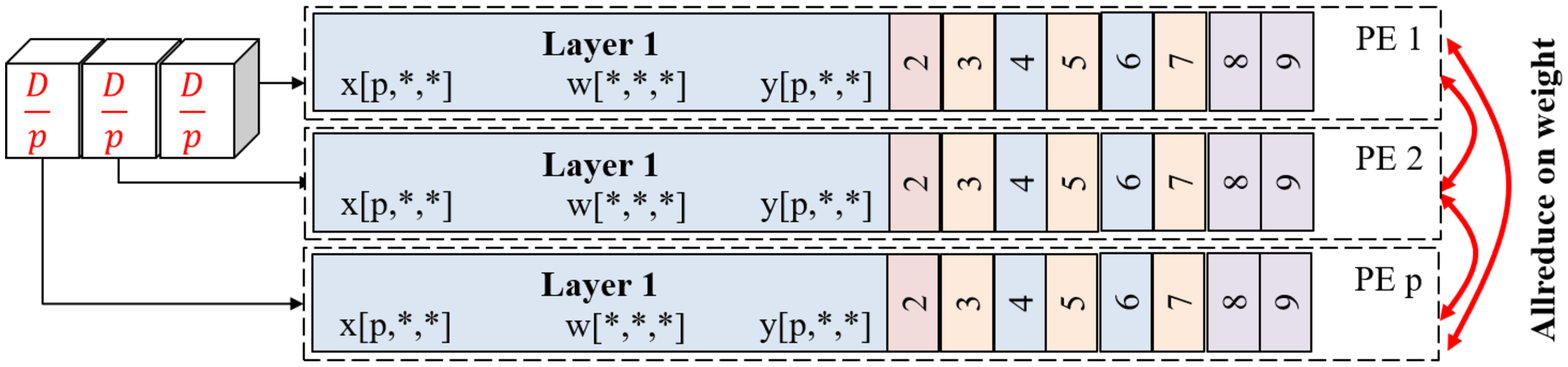}
	}
	\\
	\vspace{-0.2cm}
	\subfigure [Spatial parallelism splits the input $x$ and output $y$ on either width (as shown in this figure), height or both dimensions] {
		\label{fig:smodel}
		\includegraphics[width=7.5cm,trim=1cm 0.2cm 1cm 0cm]{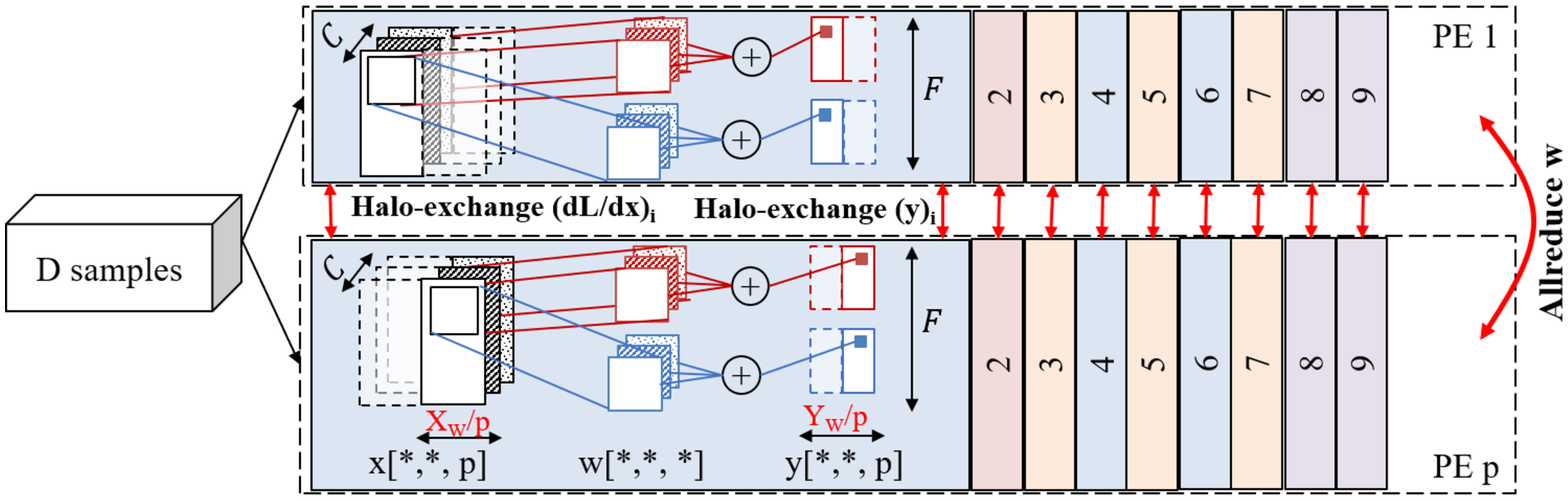}
	}
	\\
	\vspace{-0.2cm}
	\subfigure [Layer parallelism (partition the model vertically \nguyen{with pipeline implementation})] {
		\label{fig:pmodel}
		\includegraphics[width=7.5cm,trim=1cm 0.2cm 1cm 0.2cm]{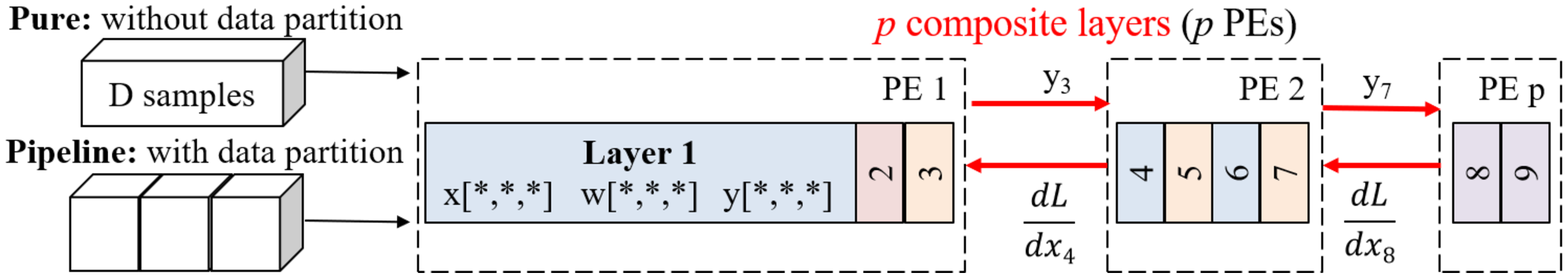}
	}
	\\
	\vspace{-0.2cm}
	\subfigure [Filter parallelism (partition the model horizontally)] {
		\label{fig:fmodel}
		\includegraphics[width=7.5cm,trim=1cm 0.2cm 1cm 0.2cm]{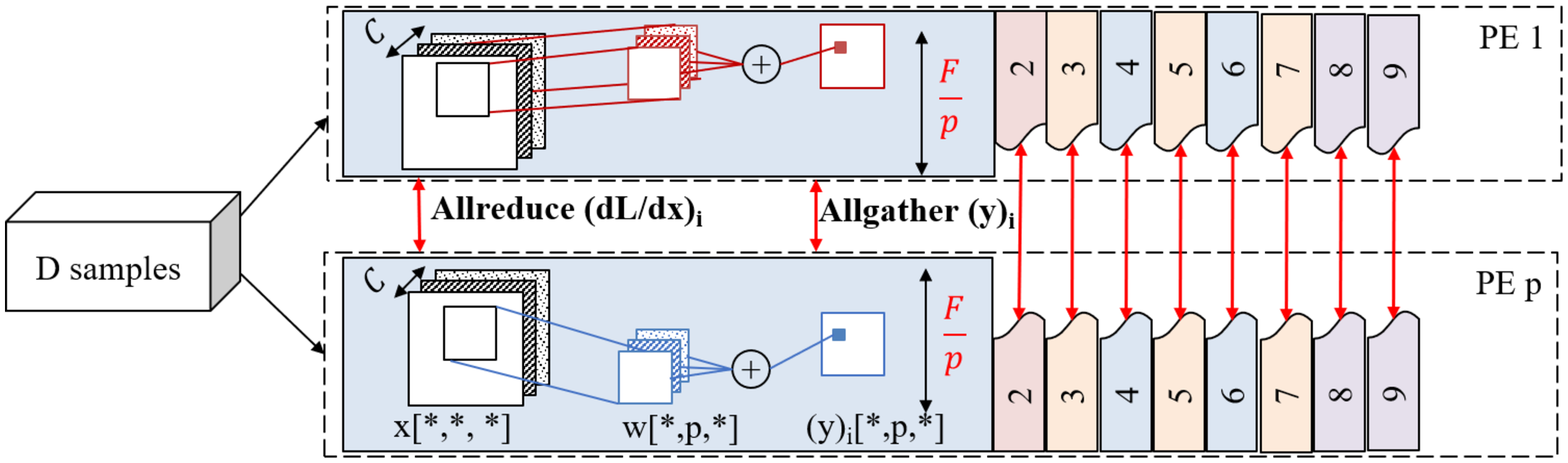}
	}
	\\
	\vspace{-0.2cm}
	\subfigure [Channel parallelism (partition the model horizontally)] {
		\label{fig:cmodel}
		\includegraphics[width=7.5cm,trim=1cm 0.2cm 1cm 0.2cm]{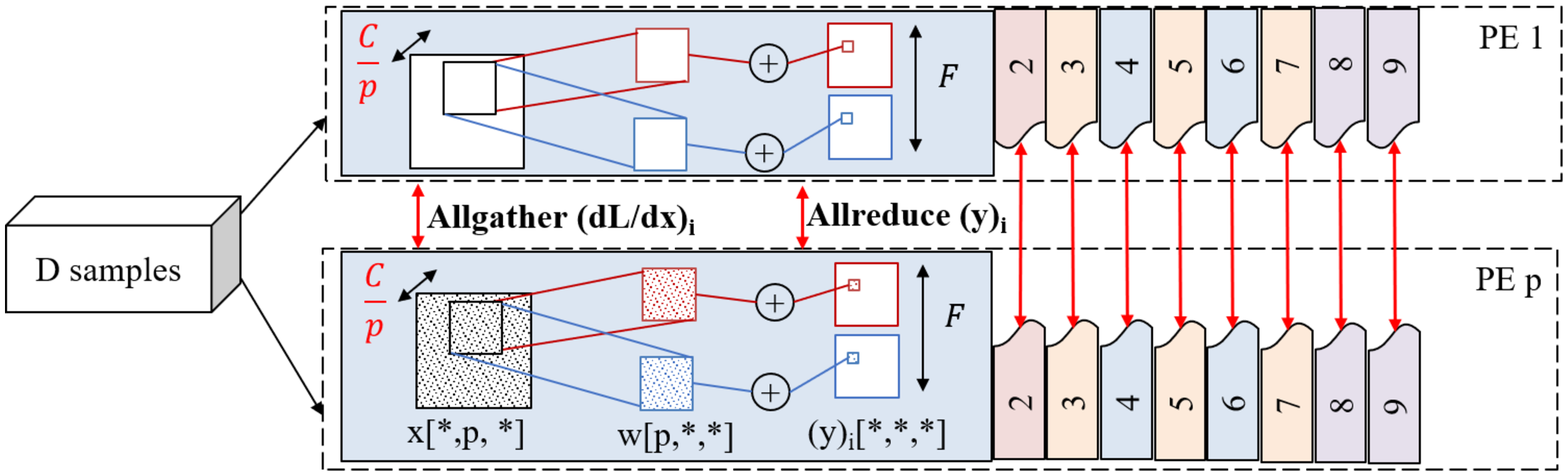}
	}
	\\
	\vspace{-0.1cm}
	\subfigure [Hybrid parallelism (example of filter on top of data parallelism)] {
		\label{fig:hmodel}
		\includegraphics[width=7.5cm,trim=1cm 0.2cm 1cm 0.2cm]{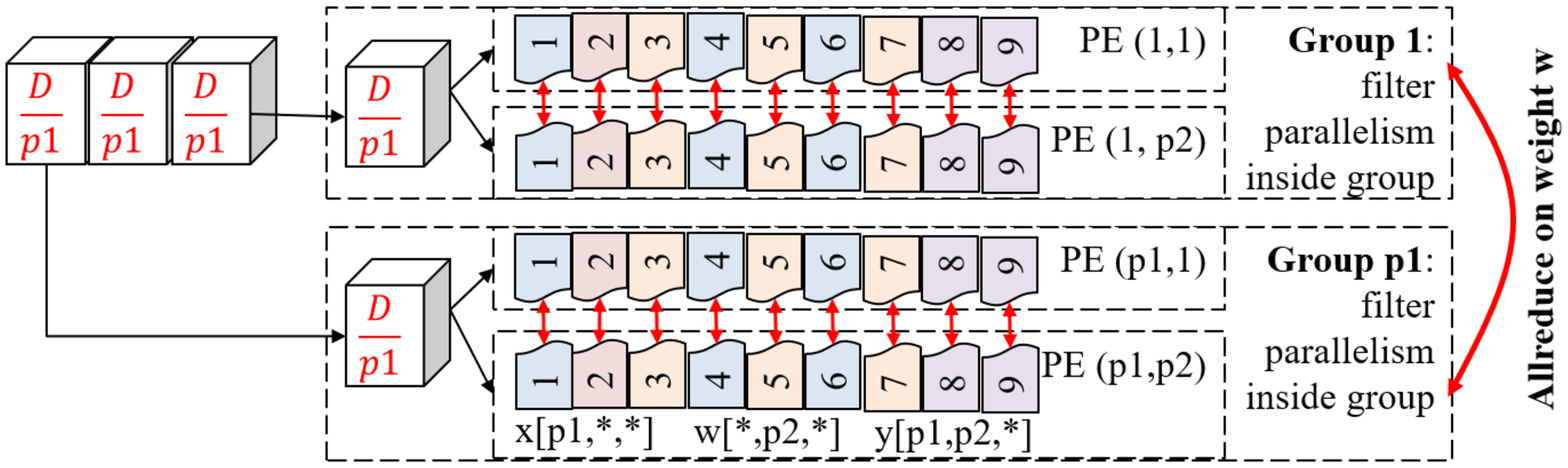}
	}
	\caption{Different strategies for distributed training of CNNs. Red solid lines refer to communication 
	}
	\label{fig:model}
	\vspace{-0.2cm}
\end{figure}

\subsection{\textbf{Data parallelism}}
%NOTE:Constrain: $p \le B$\\
The entire model is replicated on $p$ different PEs, e.g., GPUs (Figure~\ref{fig:data}) and the dataset is scattered into sub-datasets to each PE. Then the forward and backward phases are computed independently, using those different partitions of the dataset, i.e., in a micro-batch $B'=\frac{B}{p}$ at each iteration. In the gradient exchange phase, an Allreduce operation is required to aggregate the weight gradients, i.e., $\sum_{i = 1}^{p}\Big(\frac{dL}{dw}\Big)_i$. 
We define operations at the processing element $i$ in data parallelism as:

\begin{itemize}[]
	\small
	%\item[(P1)] \revision{$\text{sub-dataset}_i \xleftarrow[]{Scatter} \text{dataset},p$}.
	\item[(IO)]	$(x)_i[p,*,*] \gets IO(\text{sub-dataset}_i,B')$  in the first layer.
	\item[(FB)] $(y)_i[p,*,*] \gets FW(x_i[p,*,*], w[*,*,*])$
	\item[(FB)] $(\frac{dL}{dx})_i[p,*,*] \gets BW_{data}((\frac{dL}{dy})_i[p,*,*], w[*,*,*])$
	\item[(FB)] $(\frac{dL}{dw})_i[*,*,*]\gets BW_{weight}((\frac{dL}{dy})_i[p,*,*], (x)_i[p,*,*])$
	\item[(GE)] $\frac{dL}{dw}[*,*,*] \xleftarrow[]{Allreduce} \sum_{i=1}^{p}\big((\frac{dL}{dw})_i[*,*,*]\big)$
	\item[(WU)] $w[*,*,*] \gets WU(\frac{dL}{dw}[*,*,*])$
\end{itemize}

\subsection{\textbf{Spatial parallelism (height-width-depth)}}
All the PEs work on the same batch of samples. First, one leader PE loads those samples at each iteration and then distributes to other PEs. Note that,
the spatial dimension $H$, $W$ (and $D$ as in 3-D convolution layer), of $x$, $y$, $\frac{dL}{dx}$ and $\frac{dL}{dy}$ are split among $p$ PEs (Figure~\ref{fig:smodel}). That is $p = pw \times ph \times pd$ where $pw$, $ph$, $pd$ $\le W$, $H$, $D$, respectively. 	
Each process thus performs the forward and backward operation locally.
For a convolution layer, %and pooling layer, 
when a filter of size $K \times K$ where $K > 1$ is placed near the border of a partition, each PE requires remote data for computing. 
Thus, a small number (e.g., $\frac{K}{2}$) of rows and/or columns will be transferred from logically-neighboring remote PEs (halo exchange)~\cite{dryden2019improving}.
The exchanged data size (i.e., $\text{halo}(x_l)$) depends on how each spatial dimension is split, and the stride length.
For example, a processing element $i$ needs a halo exchange for its partial input $(x)_i$ to get $(x)_{i+}$ when computing the output $(y)_i$ in the forward phase. 
In the backward phase, the computation of $(\frac{dL}{dx})_i$ requires a halo exchange on the corresponding $(\frac{dL}{dy})_{i}$.
To compute the weight gradients requires the $(x)_{i+}$, yet no more halo exchange is required since the exchanged values of $(x)_i$ can be reused. 
In the weight update phase an Allreduce is performed for the sum of $\frac{dL}{dw}$.

\begin{itemize}[]
	\small
	\item[(IO)]	$x[*,*,*] \gets IO(\text{dataset},B)$  
	\item[(IO)]	$(x)_i[*,*,p] \xleftarrow[]{Scatter} x[*,*,*]$	in the first layer.
	\item[(FB)] $(x)_{i+}[*,*,p] \xleftarrow[]{halo} (x)_i[*,*,p]$ 
	\item[(FB)] $(y)_i[*,*,p] \gets FW((x)_{i+}[*,*,p], w[*,*,*])$
	\item[(FB)] $(\frac{dL}{dy})_{i+}[*,*,p] \xleftarrow[]{halo} (\frac{dL}{dy})_i[*,*,p]$
	\item[(FB)] $(\frac{dL}{dx})_i[*,*,p] \gets BW_{data}((\frac{dL}{dy})_{i+}[*,*,p], w[*,*,*])$
	\item[(FB)] $(\frac{dL}{dw})_i[*,*,*]\gets BW_{weight}((\frac{dL}{dy})_i[*,*,p], (x)_{i+}[*,*,p])$
	\item[(GE)] $\frac{dL}{dw}[*,*,*] \xleftarrow[]{Allreduce} \sum_{i=1}^{p}\big((\frac{dL}{dw})_i[*,*,*]\big)$
	\item[(WU)] $w[*,*,*] \gets WU(\frac{dL}{dw}[*,*,*])$
\end{itemize}

\subsection{Model-horizontal parallelism}

A model parallel variant in which each layer of the neural network model is equally divided by the number of output (filters $F$) or input channels (channels $C$) and distributed on $p$ PEs. Each PE keeps a portion of the weights of a given layer and partially computes the output in both the forward and backward phases. For example, the filter parallelism of a convolution layer~\cite{gholami2017integrated} is illustrated in Figure~\ref{fig:fmodel}. Each PE $i$ keeps $\frac{F}{p}$ filters and computes $\frac{F}{p}$ corresponding channels of the output activation. 
That is, $|(y)_i| = N \times |Y| \times \frac{F}{p}$.
After finishing the forward computation of each layer, the PEs have to share their local output, i.e., $y =  \bigcup_{i=1}^{p}(y)_{i}$ (via an Allgather operation).
After finishing the backward computation of each layer, the processes also have to share their gradient of the input (pass it to the preceding layer), i.e., $\frac{dL}{dx} = \sum_{i=1}^{p} (\frac{dL}{dx})_i$ (an Allreduce operation\footnote{In the backward phase, because a given layer $l-1$ only requires to use one partition of the layer $l$'s input gradients, i.e.,$\frac{dL}{dx}[*,p,*]$, it is possible to perform a Reduce-Scatter instead of an Allreduce operation~\cite{Naoya:sc19}.}). Because each PE performs the weight-update on its portion of weights, the gradient-exchange phase is skipped.

\begin{itemize}[]
	\small
	\item[(IO)]	$x[*,*,*] \gets IO(\text{dataset},B)$  
	\item[(IO)]	$(x)_i[*,*,*] \xleftarrow[]{Bcast} x[*,*,*]$	in the first layer.
	\item[(FB)] $(y)_i[*,p,*] \gets FW((x)_i[*,*,*], w[*,p,*])$
	\item[(FB)] $y[*,*,*] \xleftarrow[]{Allgather} \bigcup_{i=1}^{p}\big((y)_i[*,p,*]\big)$
	\item[(FB)] $(\frac{dL}{dx})_i[*,*,*] \gets BW_{data}((\frac{dL}{dy})_i[*,p,*], w[*,p,*])$	
	\item[(FB)] $\frac{dL}{dx}[*,*,*] \xleftarrow[]{Allreduce} \sum_{i=1}^{p}\big(\frac{dL}{dx})_i[*,*,*])\big)$ 
	\item[(FB)] $\frac{dL}{dw}[*,p,*]\gets BW_{weight}((\frac{dL}{dy})_i[*,p,*], (x)_i[*,*,*])$
	\item[(WU)] $w[*,p,*] \gets WU(\frac{dL}{dw}[*,p,*])$
\end{itemize}

Channel parallelism~\cite{Naoya:sc19} (Figure~\ref{fig:cmodel}) is similar to filter parallel strategy but it requires an Allreduce in the forward pass and Allgather in the backward pass.

\begin{itemize}[]
	\small
	\item[(FB)] $(y)_i[*,*,*] \gets FW((x)_i[*,p,*], w[p,*,*])$	
	\item[(FB)] $y[*,*,*] \xleftarrow[]{Allreduce}\sum_{i=1}^{p}\big((y)_i[*,*,*]\big)$	%or\\
	%$y[*,p,*] \xleftarrow[]{Reduce-Scatter}\sum_{i=1}^{p}\big((y)_i[*,*,*]\big)$
	\item[(FB)] $(\frac{dL}{dx})_i[*,p,*] \gets BW_{data}((\frac{dL}{dy})_i[*,*,*], w[p,*,*])$	
	\item[(FB)] $\frac{dL}{dx}[*,*,*] \xleftarrow[]{Allgather} \bigcup_{i=1}^{p}\big((\frac{dL}{dx})_i[*,p,*]\big)$
\end{itemize}

\subsection{\textbf{Model-vertical (layer) parallelism}}
%NOTE:
% Constrain: $p \le G$.
A model parallel variant at which the CNN is partitioned across its depth (number of layers $G$) into $p \le G$ composite layers, where each composite layer is assigned into one PE, as shown in Figure~\ref{fig:pmodel}.
We consider the pipeline implementation of this model parallelism (first proposed by GPipe~\cite{huang2018gpipe}).  
The mini-batch is divided into $S$ segments of size $\frac{B}{S}$.  
In each stage, the forward computation of a composite layer $i$-th on a data segment $s$ is performed simultaneously with the computation of composite layer $(i+1)$-th on the data segment $s-1$ and so on. 
The backward computation is done in reversed order.  

\subsection{\textbf{Hybrid parallelism}}
We have defined four different main parallel strategies which split the dimension $N$, $W \times H$ ($\times D$), $F$, $C$, and $G$, respectively. 
Without loss of generalization, a layer also can be split by the size of kernel $K \times K$. However, in practice $K$ is so small that parallelizing by dividing $K$ would not give any benefit.
Therefore, we focus on the mentioned main strategies.
A hybrid parallelism is a combination of two (or more) strategies. 
For example, Figure~\ref{fig:hmodel} illustrates the data+filter parallelism.
In which, $p$ PEs are arranged into $p1$ groups of size $p2 = \frac{p}{p1}$. 
This hybrid strategy implements the filter parallelism inside each group and data parallelism between groups. 
For a PE $1 \le i \le p2$ in a group $1 \le j \le p1$:
\begin{itemize}[]
	\small
	\item[(IO)]	$(x)_j[p1,*,*] \gets IO(\text{sub-dataset}_j,B')$
	\item[(IO)]	$(x)_{ij}[p1,*,*] \xleftarrow[]{Bcast} (x)_j[p1,*,*]$	in the first layer\\
	Filter parallelism inside a group of $p2$ PEs:
	\item[(FB)] $(y)_{ij}[p1,p2,*] \gets FW((x)_{ij}[p1,*,*], w[*,p2,*])$
	\item[(FB)] $y_{j}[p1,*,*] \xleftarrow[]{Allgather} \bigcup_{i=1}^{p2}\big((y)_{ij}[p1,p2,*]\big)$
	\item[(FB)] $(\frac{dL}{dx})_{ij}[p1,*,*] \gets BW_{data}((\frac{dL}{dy})_{ij}[p1,p2,*], w[*,p2,*])$		
	\item[(FB)] $(\frac{dL}{dx})_j[p1,*,*] \xleftarrow[]{Allreduce} \sum_{i=1}^{p2}\big(\frac{dL}{dx})_{ij}[p1,*,*])\big)$ \\
	%$\frac{dL}{dx}[p1,p2,*] \xleftarrow[]{Reduce-Scatter} \sum_{i=1}^{p2}\big(\frac{dL}{dx})_i[p1,*,*])\big)$ \\
	Data parallelism between $p1$ groups  :
	\item[(FB)] $(\frac{dL}{dw})_j[*,p2,*]\gets BW_{weight}((\frac{dL}{dy})_{ij}[p1,p2,*], (x)_{ij}[p1,*,*])$
	\item[(GE)] $\frac{dL}{dw}[*,p2,*] \xleftarrow[]{Allreduce} \sum_{j=1}^{p1}\big((\frac{dL}{dw})_j[*,p2,*]\big)$
	\item[(WU)] $w[*,p2,*] \gets WU(\frac{dL}{dw}[*,p2,*])$
\end{itemize}

Another example of hybrid parallelism is the combination of data and spatial or channel parallelism~\cite{Naoya:sc19}. Furthermore, the hybrid strategy could be more complex when applying different parallel strategies for different layers~\cite{krizhevsky2014one, jia2018exploring}.

\section{Performance Projection of Different Parallel Strategies}
%\section{A Guide for Parallel Training: Understanding the right Strategy}
\label{sec:model}
\begin{figure}[t]
		\centering
		\includegraphics[height=4.0cm,trim=1cm 0cm 1cm 0cm]{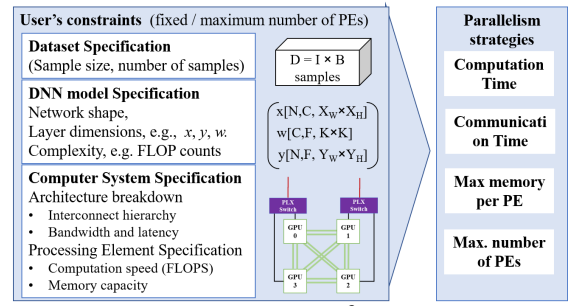}
		\caption{Overview of ParaDL}
		\label{fig:ParaDLmain}
		\vspace{-0.4cm}
\end{figure}
\subsection{Overview of ParaDL}
In this section we introduce our oracle (\emph{ParaDL}). Through the information that we can get beforehand, such as the dataset, model, supercomputer/cluster system specification, and user's constraints (e.g., maximum number of involved PEs), ParaDL calculates the computation and communication time to project the overall performance (as described in Figure~\ref{fig:ParaDLmain}).
If the strategy differs as the number of nodes increases, ParaDL would breakdown the execution time of different strategies as the number of PEs changes, i.e. scaling the number of PEs. ParaDL can be used for the following purposes:
\begin{itemize}[leftmargin=3mm]
\item Suggesting the best strategy for a given CNN, dataset, and resource budget (especially when data parallelism is not feasible).
\item Identifying the time and resources to provision from a system (we partially relied on ParaDL in this paper for that purpose when conducting our empirical experiments in Section~\ref{sec:evaluation}).
\item Comparison of projections with measured results to detect abnormal behavior (we relied on ParaDL for this purpose in our analysis of network contention in Section~\ref{sec:analytic_model}).
\item Identifying limitations of parallel strategies, shortcomings of frameworks, and bottlenecks in systems (we relied on ParaDL for this purpose in our discussion in Section~\ref{sec:result_limit}).
\item As an education tool of the parallel strategies that would improve the understanding of parallelism in DL
\end{itemize}

Frameworks that are used for DL are comprised of complex and interleaved layers of optimized functions. A pure analytical model of parallel strategies in CNNs would, therefore, be impractical. In this paper we adopt a hybrid analytical/empirical modeling approach at which we: (i) use analytical modeling for functional requirements driven by the parallelism strategies (Section~\ref{sec:analytic_model}), and (ii) empirical parametrization for functions not related to the parallel strategy being deployed (more details in Section~\ref{sec:paradl:perf}). Finally, we quantify the accuracy of the oracle with a large empirical evaluation in Section~\ref{sec:evaluation}.

\subsection{Assumptions and Restrictions}
The study in this paper is based on the following assumptions.
\label{sec:assumption}

\textbf{Targeted models and datasets:} our study covers all types of layers used in production CNNs, and could hence be used for projecting the performance of any production CNN model, not just the models we evaluate in the paper. We also support the input (i.e. samples) to be of any dimension (as shown in Table~\ref{table:notation}). 

\textbf{Training time and memory estimation:} Our study focuses predominantly on the computation and communication time of the CNN training, thus we assume that all the training data is available in memory before starting the training process. In other words, in this model we do not include the time for I/O. 

One could conservatively estimate the memory required on a per layer basis by assuming the memory buffers of the output of layer $l$ are different from the memory buffers for the input of layer $l+1$, however, in reality both buffers being the same. Additionally, in reality there is a variety of optimizations that frameworks implement to reduce the memory used (See Table~\ref{table:relatedwork}).  
Since those optimization methods are complexly intertwined and depend on the framework implementations, without loss of generalization, we propose a practical memory requirement estimation.  More specifically, we start out from the naive memory projection that aggregates layers, then we reduce that conservative upper bound to reflect the actual memory optimizations happening inside frameworks. We introduce a memory reuse factor $\gamma$. The actual minimum required memory, after all memory reuse optimizations are applied, can be estimated by multiplying total naive required memory by $\gamma$. This memory reuse factor can be derived from several elaborate studies on model-level and layer-level memory profiling of CNNs~\cite{Li2019BenanzaA,10.1145/3243904}.

\textbf{Parallel strategies:} all results in this paper, unless otherwise stated, are for the de facto scaling approach in DL: weak scaling. The mini-batch size scales with the number of PE, hence the number of samples per PE remains constant. 
In addition, unless mentioned, we do not actively optimize for changing the type of parallelism between different layers in a model, i.e., different layers do not have different parallel strategy. However, there can be cases at which a different type of parallelism is used, in order to avoid performance degradation. For instance, the fully connected layer in spatial parallelism is not spatially parallelized, since that would incur high communication overhead for a layer that is typically a fraction of the compute cost of convolution layers~\cite{krizhevsky2014one}.

\subsection{Performance and Memory Projection}
\label{sec:analytic_model}
\renewcommand{\arraystretch}{1.0}
\begin{table*}[]
	\centering
	\scriptsize
	\caption{Computation, Communication, and Memory Analysis Summary (per epoch)}
%	\begin{tabular}{|p{0.9cm}|l|l|l|p{2.2cm}|}
		\resizebox{\textwidth}{!}{\begin{tabular}{|l|l|l|l|l|}
		\hline
		& \textbf{Computation Time $T_{comp}$} & \textbf{Communication Time $T_{comm}$} & \textbf{Maximum Memory Per PE} & \textbf{Number of PEs $p$}\\ 
		\hline
		\textbf{Serial} & 
		$D\sum_{l= 1}^{G}\Big(FW_l + BW_l\Big) + \frac{D}{B}\sum_{l = 1}^{G}(WU_l)$& 
		0 & 
		$\gamma\delta\sum_{l=1}^{G}\Big(2B(|x_l| + |y_l|) + 2|w_l| +\nguyen{|bi_l|} \Big)$ &
		$p = 1$\\ 
		\hline
		\textbf{Data} & 
		$\frac{D}{p}\sum_{l = 1}^{G}(FW_l + BW_l) + \frac{D}{B}\sum_{l = 1}^{G}(WU_l)$ &  
		$2\frac{D}{B}(p-1)\Big(\alpha + \frac{\sum_{l=1}^{G}|w_l|}{p}\delta\beta\Big)$ & 
		$\gamma\delta\sum_{l=1}^{G}\Big(\frac{2B}{p}(|x_l| + |y_l|) + 2|w_l| + \nguyen{|bi_l|}\Big)$ &
		$p \le B$\\ 
		\hline
		\textbf{Spatial} &
		$ \frac{D}{p}\sum_{l = 1}^{G}\Big(FW_l + BW_l\Big) + \frac{D}{B}\sum_{l = 1}^{G}(WU_l)$  &
		\begin{tabular}[c]{@{}l@{}}	$2\frac{D}{B}\Big((p-1)(\alpha + \frac{\sum_{l=1}^{G}|w_l|}{p}\delta\beta) + $\\ 
			%$\frac{D}{B}
			$\sum_{l=1}^{G}\big(2\alpha + B(\text{halo}(|x_l|) + \text{halo}(|\frac{dL}{dy_l}|))\delta\beta \big)\Big)$ \end{tabular} &
		$\gamma\delta\sum_{l=1}^{G}\Big(2B\frac{(|x_l| + |y_l|)}{p} + 2|w_l| + \nguyen{|bi_l|}\Big)$  &
		\begin{tabular}[c]{@{}l@{}}	$p = pw \times ph \le$\\ $\min_{l=1}^{G}(W_l \times H_l)$ \end{tabular}\\
		\hline
		\begin{tabular}[c]{@{}l@{}}\textbf{Layer} \\(Pipeline)\end{tabular}&
		\begin{tabular}[c]{@{}l@{}}
			$\frac{D(p + S-1)}{S}\Big(\max_{i = 1}^{p}(FW_{G_i})$ \\ $+\max_{i = 1}^{p}(BW_{G_i})\Big) + \max_{i = 1}^{p}(WU_{G_i})\Big) $ \end{tabular}&
		%$\frac{(p + S-1)D}{Sp}\sum_{l = 1}^{G}\Big(FW_l + BW_l\Big) $ &
		%$(p + S - 2)\Big( \max_{i =1}^{p}\big(\alpha + \frac{B}{S}|y_{G_i}|\delta\beta\big) + \max_{i =1}^{p}\big(\alpha + \frac{B}{S}|\frac{dL}{dx_{G_i}}|\delta\beta\big)\Big)$
		$2\frac{D(p + S - 2)}{B}\Big( \max_{i =1}^{p-1}\big(\alpha + \frac{B}{S}|y_{G_i}|\delta\beta\big)\Big)$ &
		\begin{tabular}[c]{@{}l@{}}$\gamma\delta\max_{i=1}^{p}\Big(\sum_{l=1}^{G_i}\big(2B(|x_l| + |y_l|)$ \\ 
			$ +2|w_l| + \nguyen{|bi_l|}\big)\Big)$ \end{tabular} &
		$p \le G$\\  
		\hline
		\textbf{Filter} & 
		$\frac{D}{p}\sum_{l = 1}^{G}\Big(FW_l + BW_l\Big) + \frac{D}{Bp}\sum_{l = 1}^{G}(WU_l)$&  
		$3\frac{D}{B}(p-1)\sum_{l=1}^{G-1}(\alpha + \frac{B|y_l|}{p}\delta\beta)$& 
		%$3\frac{D}{B}(p-1)\Big((G-1)\alpha + \frac{B\sum_{l=1}^{G-1}|y_l|}{p}\delta\beta\Big)$&
		$\gamma\delta\sum_{l=1}^{G}\Big(2B(|x_l| + |y_l|) + \frac{2|w_l|}{p} + \nguyen{|bi_l|}\Big)$ &
		$p \le \min_{l=1}^{G}(F_l)$\\  
		\hline
		\textbf{Channel} & 
		$\frac{D}{p}\sum_{l = 1}^{G}\Big(FW_l + BW_l\Big) + \frac{D}{Bp}\sum_{l = 1}^{G}(WU_l)$&  
		$3\frac{D}{B}(p-1)\sum_{l=1}^{G-1}(\alpha + \frac{B|y_l|}{p}\delta\beta)$ &
		%$3\frac{D}{B}(p-1)\Big((G-1)\alpha + \frac{B\sum_{l=1}^{G-1}|y_l|}{p}\delta\beta\Big)$&
		$\gamma\delta\sum_{l=1}^{G}\Big(2B(|x_l| + |y_l|) + \frac{2|w_l|}{p} + \nguyen{|bi_l|}\Big)$ &
		$p \le \min_{l=1}^{G}(C_l)$\\  
		\hline
		\begin{tabular}[c]{@{}l@{}}\textbf{Data} + \\\textbf{Filter}\end{tabular}&
		$\frac{D}{p}\sum_{l = 1}^{G}\Big(FW_l + BW_l\Big) + \frac{D}{Bp2}\sum_{l = 1}^{G}(WU_l)$ &
		\begin{tabular}[c]{@{}l@{}} $3\frac{D}{B}(p2-1)\sum_{l=1}^{G-1}(\alpha + \frac{B|y_l|}{p}\delta\beta) +$ \\$2\frac{D}{B}(p1-1)(\alpha + \frac{\sum_{l=1}^{G}|w_l|}{p}\delta\beta)$ \end{tabular}&
		$\gamma\delta\sum_{l=1}^{G}\Big(\frac{2B(|x_l| + |y_l|)}{p1} + \frac{2|w_l|}{p2} + \nguyen{|bi_l|}\Big)$&
		\begin{tabular}[c]{@{}l@{}}$p = p1 \times p2 \le $ \\ $B \times \min_{l=1}^{G}(F_l)$\end{tabular} \\   
		\hline
	\end{tabular}}
	\label{table:sum_time}
	\vspace{-0.3cm}
\end{table*}
\renewcommand{\arraystretch}{1.0}

In this section, we estimate the total training time in one epoch and maximum memory per PE for the mentioned main parallel strategies, including hybrid. Let $FW_l$, $BW_l$ denote the time to perform the computation of forward and backward propagation for one sample and let $WU_l$ denote the time for weight update per iteration at layer $l$~\footnote{In pipeline, each PE $i$ keeps $G_i$ layers of the model given that $\sum_{i=1}^{p}G_i = G$.
Let $FW_{G_i}$, $BW_{G_i}$ and $WU_{G_i}$ denote the time for performing the forward,  backward, and weight update computation of group $i$ per sample.}. 
$T_{ar}(p,m)$, $T_{ag}(p,m)$, and $T_{p2p}(m)$ stand for the time of transferring a data buffer of $m$-size between $p$ PEs via an Allreduce, Allgather, and a peer-to-peer scheme, respectively.
In data parallelism, the training includes both computation and communication time.
Each PE processes a micro batch size $B' = \frac{B}{p}$ in this case. The time for FW and BW in one iteration is $\frac{1}{p}$ of the single-process.
Thus the total computation time in one epoch becomes:
\setlength{\abovedisplayskip}{1pt}
\setlength{\belowdisplayskip}{1pt}
\begin{equation}
\label{equal:data_comp_2}
\scriptsize
\begin{split}
T_{data,comp} %= \sum_{1}^{I}\sum_{l = 1}^{G}(\frac{B}{p}(FW_l + BW_l) + WU_l) 
= \frac{D}{p}\sum_{l = 1}^{G}(FW_l + BW_l) + \frac{D}{B}\sum_{l = 1}^{G}(WU_l)
\end{split}
\end{equation}

Because PEs have to share their gradients at the end of each iteration, the time for communication is $\frac{D}{B}T_{ar}(p,\sum_{l=1}^{G}|w_l|)$.
Considering the memory footprint, in data parallelism we duplicate the entire model on $p$ different PEs. Each PE processes a partition of the dataset in a microbatch of $B'=\frac{B}{p}$ samples.
A layer $l$ needs memory to store its input $B'|x_l|$, activation $B'|y_l|$, weights $|w_l|$, \nguyen{bias $|bi_l|$}, the gradients $B'|\frac{dL}{dx_l}|$, $B'|\frac{dL}{dy_l}|$, and $|\frac{dL}{dw_l}|$. 
Overall, if each item of the input, activation, weight and gradients are stored in $\delta$ bytes, 
the maximum required memory at one PE is:
\begin{equation}
\label{equal:data_mem_1}
\scriptsize
\begin{split}
M_{data} = \sum_{l=1}^{G}\delta(B'(|x_l|+|y_l|)+ |w_l| + \nguyen{|bi_l|} +  B'(|\frac{dL}{dx_l}|+|\frac{dL}{dy_l}|) + |\frac{dL}{dw_l}|)
%\\ 
%\revision{=  2\delta\sum_{l=1}^{G}\Big(\frac{B}{p}(|x_l| + |y_l|) + |w_l| \Big)}
\end{split}
\end{equation} 

In theory, the computation time can be estimated by observing the dataset and CNN model (e.g., FLOP counts and the computation speed of each PE).
For modeling the communication time, there exist various derived / specific analytical performance models, e.g., as in the survey of Rico Gallego et. al.~\cite{Rico_comm_model}. To keep the performance modeling generic, we choose to use the Hockney $\alpha-\beta$ model, in which, the peer-to-peer communication time of transferring a message of size $m$ is modeled by $T_{p2p}(p,m)=\alpha + m\beta$.
Time for a message send from a source to a destination is $\alpha$ (also known as startup time) and  the time to inject one byte of data into the network is $\beta$.
We follow the common practice in DL communication libraries such as NCCL~\cite{nccl} to use a ring-based algorithm for all the collective communication operation with large message sizes and a tree-based algorithm for small message sizes.
In the ring-based algorithm, a logical ring is first constructed among $p$ PEs based on the system network architecture.
Then, each PE partitions its $m$-size data buffer into $p$ segments of size $\frac{m}{p}$.
Each PE then sends one data segment to the successive PE and receive another segment from the preceding PE along the ring, i.e., a total of $p-1$ steps for Allgather and $2(p-1)$ steps for Allreduce.
Thus $T_{ar}(p,m)$ and $T_{ag}(p,m)$ can be modeled by $2(p-1)(\alpha + \frac{m}{p}\beta)$ and $(p-1)(\alpha + m\beta)$, respectively.
Based on this communication model, %we summarize our analysis of parallel strategies in Table~\ref{table:sum_time}.
we also estimate the total training time in one epoch and the maximum memory required per PE for the mentioned parallel strategies.
We summarize our analytical model in Table~\ref{table:sum_time}~\footnote{\nguyen{We note that the proposed communication model can naturally be extended for different Allreduce schemes or algorithms, such as Parameter Server or tree-based algorithms. For example,} when message sizes are small, communication time with tree-based algorithms can be estimated as $2(\log(p)+k)(\alpha + m\frac{m}{2k}\beta)$ where a message is divided into $k$ chunks to communicate in a pipeline~\cite{SANDERS2009581}.}. The details of this analysis can be found in the Appendix of this paper.

\textbf{Contention modeling}: Ideally, in a system without contention, the start up time $\alpha$ of a given pair is estimated as the total switching latency, which depends on the number of intermediate switching elements. In addition, $\beta$ is the inverse of the minimum link bandwidth on the routing path between two PEs (the bottleneck link). However, network congestion is one of the biggest problems facing HPC systems today, affecting system throughput and performance. To address the contention effects we introduce the use of a contention penalty coefficient $\phi$, which divides the bandwidth of a link by the number of communication flows $\phi$ sharing this link at each step of collective communications~\cite{KIM20011692_Contention}. 
In our analytical model, we only consider the self-contention caused by all the communication flows of the training process itself, e.g., a link is shared between different groups in hybrid parallelism strategies. The contention coefficient can be estimated analytically by using dynamic contention graphs~\cite{Martomasso_Contention}. It is important to note that we do not intend to model the contention caused by congested networks due to a large number of applications running at the same time in a shared system. Such kind of external contention affects all parallelism strategies and do not reflect the baseline fundamental performance of each parallelism strategy. In addition, the baseline performance predicted by our analytical model can be complemented with a congestion impact factor, which can be empirically estimated as in~\cite{GPCNeT}, in order to predict the real-world performance in production environments.

\subsection{Empirical Parametrization}
\label{sec:paradl:perf}
As mentioned earlier, we rely on a hybrid of analytical modeling and empirical parametrization for ParaDL. To reduce the impact of noise associated with black-box empirical modeling~\cite{copik2020extracting}, we segment the experiments used to inspect the target parameters. We are thus able to distinguish between effects of noise on the measurements and actual runtime change because of parameter influence. The empirical parameters are (as defined in Table~\ref{table:notation}):
\begin{itemize}[leftmargin=3mm]
    \item \textbf{Computation parameters} ($FW_l$, $BW_l$, and $WU_l$): It is important to note that processors, CPUs and GPUs, rarely perform close to their peak performance. We empirically profile the average computation time per sample of each layer (or group of layers) on the target architecture to get a more accurate result. Such profiling can be performed easily and quickly beforehand. Furthermore, the empirical compute time, per a given layer on a given processor, is available in DL databases of models~\cite{Li2019BenanzaA}.
    \item \textbf{Communication parameters} ($\alpha$ and $\beta$): The interconnect hierarchy of modern computing systems, the algorithms used by communication libraries, and the communication technologies (such as GPUDirect~\cite{gpudirect}) may lead to differences in the latency and bandwidth factors $\alpha$ and $\beta$. %in different configuration.
    Thus, we empirically measure the communication time of collective communication patterns, such as Allreduce, with different message size, number of involved processing elements on a specific computing system. Those empirical measurements can be derived from well-known tools for performance of systems, e.g., OSU Micro-Benchmarks or NCCL-test~\cite{AngLi2019NvSwitchEvaluate}. We then use those benchmark results to interpolate $\alpha$ and $\beta$. \nguyen{It is important to note that $\alpha$ and $\beta$ become different when changing the number of processing elements in a hierarchical computing architecture, e.g. intra-node, intra-rack and, inter-rack communication.}

\end{itemize}
It is important to emphasize that the empirical parameters in our model are invariant to the implementation of the parallelism strategies, i.e., values of empirical parameters could change when moving from one framework to another, yet values of the analytical parameters would not. Finally, to simplify the portability of ParaDL between different frameworks and systems, we include the following with the ParaDL utility: a) detailed instructions of using the benchmarks used for gathering the empirical parameters we use, and b) pointers to DL model and layers databases from which the user could get empirical breakdown of compute and memory requirement at the granularity of layers.

\subsection{Implementation}
\label{sec:implementation}

\subsubsection{Implementation Details}
\albert{
We implement data, channel, filter, spatial and hybrid parallelism strategies using ChainerMN \cite{chainermn} 
for distributed execution. Although ChainerMN provides a built-in implementation for data parallelism and some minimum level of support for model parallelism, it is not sufficient for testing all the parallel strategies we study here (the same insufficiency also goes for PyTorch, TensorFlow, and others). Substantial engineering effort was required to modify and extend the existing implementation and create \textbf{ChainerMNX} to support all forms of parallelism. This extension included modifying existing communicators meant for data parallelism to support hybrid parallelism. We also extended existing convolutional layers to support filter/channel/spatial convolutions\footnote{The code is publicly available here \url{https://github.com/ankahira/chainermnx}}. We mark our implementation for each type of targeted layers and parallel strategies in Table~\ref{table:implement_layer}. }

\begin{table}[t]
	%\small
	\center
	\scriptsize
	\caption{Implementation Overview($\checkmark$: customized; - : untouched)}
\resizebox{\linewidth}{!}{
	\begin{tabular}{|l|c|c|c|c|c|}
		\hline
		%\multicolumn{1}{|c|}{\multirow{2}{*}{\textbf{\begin{tabular}[c]{@{}c@{}}Parallelism \\ strategy\end{tabular}}}} & \multicolumn{6}{c|}{\textbf{Layers}} \\ \cline{2-7} 
		\multicolumn{1}{|c|}{\textbf{\begin{tabular}[c]{@{}c@{}}Parallelism \\ strategy\end{tabular}}} 
		%\multicolumn{1}{|c|}{} 
		& \multicolumn{1}{c|}{Conv} & \multicolumn{1}{c|}{Pooling} & \multicolumn{1}{c|}{\begin{tabular}[c]{@{}c@{}}BNorm / LNorm\end{tabular}} & \multicolumn{1}{c|}{ReLU} & \multicolumn{1}{c|}{FC}\\ \hline
		Data & - & - & - & - & - \\ \hline
		Spatial & $\checkmark$ & $\checkmark$ & - & - & $\checkmark$\\ \hline
		%Layer (pipeline) & $\checkmark$ &  $\checkmark$ & - & $\checkmark$ & $\checkmark$\\ \hline
		Filter / Channel & $\checkmark$ & -  & - & - & $\checkmark$  \\ \hline
		%Channel & $\checkmark$ & - & - & - &  $\checkmark$\\ \hline
	\end{tabular}
}
	\label{table:implement_layer}
	\vspace{-0.3cm}
\end{table}

\leo{More specifically, we use the default implementation of Chainer for data parallelism. Since the size of each dimension (i.e., $H$, $W$, and/or $D$) limits the parallelism of spatial strategy, in this work, we implement the spatial strategy for some first layers of a given model until adequate parallelism is exposed while still maintaining the maximum required memory per node within memory capacity. We then implement an Allgather to collect the full set of activations before passing it to the following layers which perform similar to the sequential implementation. For example, we aggregate after the final convolution layer (before a fully-connected layer) in VGG16, ResNet-50 and ResNet-152. For CosmoFlow, we aggregate after the second convolution/pooling layer because most of required memory footprint and compute are in those first two layers.} 

The minimum number of input channels $C$ at each layer limits the parallelism of channel strategy, e.g., only $3$ input channels for ImageNet. In this work we implement the channel parallelism from the second layer. For hybrid strategies such as Data+Filter (or Data+Spatial), we map the data parallelism inter-node. This implementation is also used by Dryden et al.~\cite{Naoya:sc19}. 
We leverage ChainerMN with MPI support for both inter-node and intra-node communication. 	
To perform an Allreduce and update the gradients in hybrid strategies, we use ChainerMN's multi-node optimizer which wraps an optimizer and performs an Allreduce before updating the gradients. 
For the Data+Spatial parallelism, we perform a reduce inside each node to a leader GPU, then perform an Allreduce between the leaders. These two Allreduce involve different parallelization techniques (i.e., spatial in local and data in global). For the Data+Filter parallelism, we perform a segmented Allreduce, e.g., disjoint subsets of GPU run Allreduces on different sets of the weights, i.e., number of subsets equals to number of GPUs per node.

\subsubsection{Accuracy and Correctness}
We aim at making sure our implementation of different parallelism strategies have the same behavior as data parallelism. We first compare the output activations/gradients (in forward/backward phases) of each layer (value-by-value) to confirm that the parallelization artifacts, e.g., halo exchange, do not affect the correctness. Note that these new implementations change only the decomposition of the tensors, and do not change any operator or hyper-paramaters that have an impact on accuracy.

Second, in this work we assure that batch normalization (BN) layers are supported in all parallel strategies since the accuracy of training can be affected by the implementation of the BN layers ~\cite{IoffeS15_BatchNorm, Zhang_2018_CVPR}.
More specifically, for the data parallelism strategy, the typical implementations of batch normalization in commonly used frameworks such as Caffe, PyTorch, TensorFlow are all unsynchronized. This implementation leads to data being only normalized within each PE, separately. \albert{In typical cases, the local batch-size is usually already large enough for BN layers to function as intended}. Yet in some cases, the local batch-size will be only $2$ or $4$ in each PE, which will lead to significant sample bias, and further degrade the accuracy. In this case, we suggest to use the synchronized BN implementation as mentioned in~\cite{Zhang_2018_CVPR}, which requires a communication overhead for computing the global mini-batch mean. For the spatial strategy, although performing batch normalization on subsets of the spatial dimensions has not been explored, to the authors knowledge, this computation requires no significant adjustment ~\cite{dryden2019improving}.
More specifically, BN is typically computed locally on each PE on its own portion of the spatially partitioned data.

In the filter and channel parallelism strategy, since all PEs keep the same set of activations after performing the Allgather operation at each layer, the BN layer could be implemented as in the sequential strategy. It could be implemented in a centralized fashion, e.g., one PE performs the BN and then sends the result to other PEs. Alternatively, each node could redundantly compute the BN layer (distributed approach).
In this work, we use the distributed approach which does not require any communication overhead.

\section{Evaluation}
\label{sec:evaluation}
\begin{figure*}
	\centering
	\includegraphics[width=\linewidth ,trim=0cm -0.2cm 0cm 0cm]{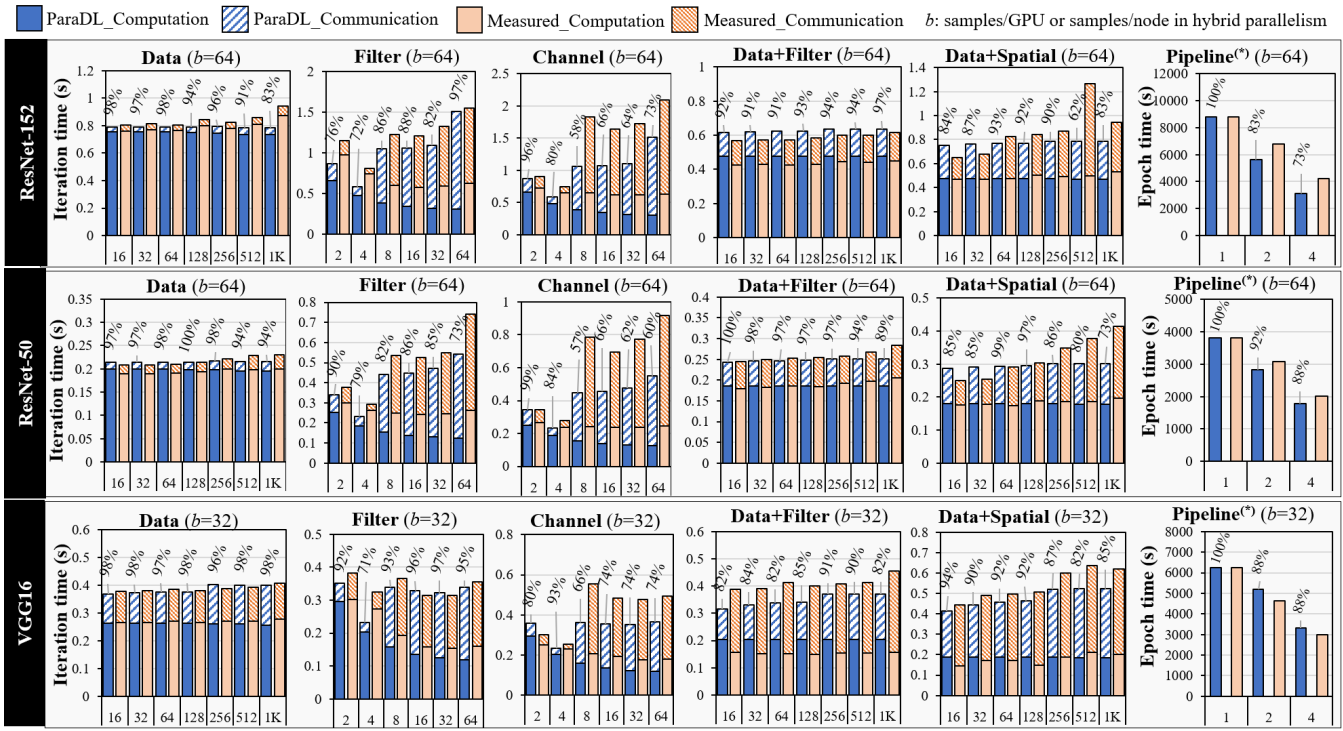}
	\caption{Time breakdown of our analytical model (ParaDL) in comparison with measured runs. The label above each column shows the \textit{projection accuracy}. The x-axis is the number of GPUs. Filter/channel are strong scaling.% Data is with weak scaling.
	\small\textsuperscript{(*)}Values are total time since pipeline parallelism~\cite{kim2020torchgpipe} overlaps the computation and communication. 
	}
	\vspace{-0.3cm}
	\label{fig:comparision}
\end{figure*}
\begin{figure}[t]
	\begin{minipage}{0.43\linewidth}
		\begin{figure}[H]
			\includegraphics[width=\linewidth]{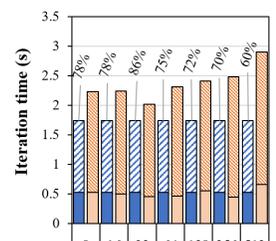}
			\caption{Prediction Accuracy of ParaDL with CosmoFlow for Data+Spatial}
			\label{fig:accuracy_cosmoflow}
		\end{figure}
	\end{minipage}
	\hspace{0.1cm}
	\begin{minipage}{0.53\linewidth}
		\begin{figure}[H]
			\includegraphics[width=\linewidth]{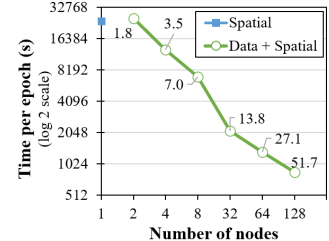}
			\caption{\small{Spatial + data scaling with CosmoFlow. The labels show the speedup ratio of spatial+data over the pure spatial strategy}} %(with 4GPUs on 1 node).}}
			\label{fig:ds_CosmoFlow}
		\end{figure}	
	\end{minipage}
	\vspace{-0.5cm}
\end{figure}

\albert{In this section, we describe how we conduct a wide range of experiments to show the accuracy and utility of ParaDL in projecting the performance} of distributed training of CNN models under different parallel strategies, including hybrid ones. We compare ParaDL projection results to the empirical measurements on a  multi-petaflop HPC system with thousands of GPUs. 
In addition, we characterize the bottlenecks and limitations of different parallelization strategies, and highlight relevant findings observed with the help of ParaDL.

\subsection{Methodology}
\label{sec:methodology}
%\subsection{Selected Models and Datasets}
%\label{sec:selected}
\textbf{Selected Models and Datasets:} We choose different CNN models and datasets with different characteristics that affect performance and memory requirements. They are summarized in Table~\ref{table:test_case}.

\begin{table}[t]
\centering
\caption{Models and Datasets Used in Experiments}%. \textit{tz}: token size.}
\scriptsize
\label{table:test_case}
\setlength\tabcolsep{3pt} % default value: 6pt
\resizebox{\linewidth}{!}{
	\begin{tabular}{|l|c|c|c|c|}
		\hline
		\multicolumn{1}{|c|}{Model} & Dataset             & \#Samples (Size)& \# Param. & \#Layers \\ \hline
		ResNet-50~\cite{he2016deep}                            &    \multirow{3}{*}{ImageNet~\cite{russakovsky2015imagenet}}     & \multirow{3}{*}{1.28M (3$\times226^2$)}& $\approx$ 25M      & 50                 \\ \cline{1-1} \cline{4-5} 
		ResNet-152~\cite{he2016deep}                           &                               &            & $\approx$ 58M      & 152  
		 \\\cline{1-1} \cline{4-5}
		VGG16~\cite{DBLP:journals/corr/SimonyanZ14a}                               &                               &            & $\approx$ 169M     & 38                 
		%\\\cline{1-1} \cline{4-5}
		%AlexNet~\cite{krizhevsky2012imagenet}                           &                               &            & $\approx$ 30M & 20 & 0.8M 
		\\ \hline
		CosmoFlow~\cite{Mathuriya:2018:CUD:3291656.3291743}      & CosmoFlow~\cite{CosmoflowDataset}     & 1584 (4$\times256^3$) & $\approx$ 2M                   & 20  \\  \hline 
		%\midrule Megatron-LM~\cite{Shoeybi2019MegatronLMTM}      & OpenWT~\cite{owt}     & 7.2M (1024$\times$ \textit{tz}) & $\approx$ 2.5B                   & 54 
		%\\  \hline
\end{tabular}}
\vspace{-0.3cm}
\end{table}
\renewcommand{\arraystretch}{1.0}
\normalsize
\textbf{Evaluation Environment:} 
%\label{sec:environment}
Experiments are performed on a multi-petaflop supercomputer, with two Intel Xeon Gold 6148 Processors and four NVIDIA Tesla V100 GPUs (16GB of memory per GPU) on each compute node. The GPUs are connected intra-node to the CPUs by PLX switches and PCIe Gen3 x16 links (16 GBps), and together by NVLink (20 GBps).  The compute nodes are connected in a 3-level fat-tree topology which has full-bisection bandwidth, and 1:3 over-subscription for intra-rack and inter-rack, respectively (two InfiniBand EDR, e.g., 12.5 GBps, per compute node and $17$ compute nodes per rack). 

{\bf Configurations of Experiments:} 
We perform the experiments of the parallel strategies using the framework Chainer~\cite{chainer} (v7.0.0), ChainerMN~\cite{chainermn} (the multi-node varient of Chainer), and CUDA (v10.0).
We also use the PyTorch (v1.5) implementation for the pipeline strategy~\cite{kim2020torchgpipe}.
%and Megatron-LM~\cite{wahib2020scaling}
We implement all communication functions based on Nvidia's NCCL library~\cite{nccl} (v2.4.8.1). The exceptions are at which we use MPI (OpenMPI v2.1.6): a) the halo exchange of the spatial strategy since P2P communication interfaces are not supported by NCCL\footnote{The latest version of NCCL now supports P2P communications.}, and b) the Allgatherv for the spatial strategy since NCCL does not support Allgatherv.

An important performance factor is efficient device utilization of GPUs. Thus, we conducted a series of test runs for each type of parallelism and DL model to identify the optimal number of samples per GPU (or node) that would efficiently utilize the device (marked as $b$ in Figure~\ref{fig:comparision}).
We observed that the performance drops significantly when we train using a higher samples/GPU number than the optimal one. This occurs when the computational load becomes too large to effectively utilize a single GPU. For CosmoFlow with spatial strategies, since we use only one sample per node, i.e., 0.25 samples/GPU: we could not have the freedom to tune the parameter $b$. This is often the case for models using large 3D input datasets (increasingly common in scientific computing), where data parallelism is simply not an option. \leo{Given that it was not possible to get the empirical layer by layer time for CosmoFlow running a sequential implementation with the $512^3$ data size, we used $256^3$ sample sizes and multiplied the computation time by 8. We confirmed with measurements that the strategy was accurate.}

\subsection{ParaDL's Projection and Accuracy}
\label{sec:accuracy}
This section discusses how close is the projection of the ParaDL oracle in comparison with the measured experiments. It is a complex task to accurately project the performance of DL training, especially when scaling. More specifically, the following factors have a significant effect on performance: contention on the PFS, the effectiveness of the pipeline used for asynchronous data loading, network contention, implementation quality, and overheads of solution fidelity book-keeping. That being said, in this section we aim to demonstrate that the presented oracle, despite the complexities mentioned above, reasonably represents the reality of measured runs on an actual system (especially when scaling up to $1024$ GPUs). In this comparison, we focus only on the computation and communication time of the main training loop \leo{(the most time consuming part)} and exclude other times from this study such as I/O staging. %(See Section~\ref{sec:staging}).

Figure~\ref{fig:comparision} shows the oracle's projections versus the measured runs for different parallel strategies using three different models (a fourth model is shown in Figure~\ref{fig:accuracy_cosmoflow}).
We ran all the permutations of possible configurations but plot only some of them because of space limitations. The figure is divided in three rows, one for each CNN model, and six columns, one for each parallelism strategy. The parameter $b$ shows the mini-batch size for each case. As mentioned in Section~\ref{sec:methodology}, the mini-batch size is set to achieve the highest device occupancy on GPUs. The x-axis shows the number of GPUs, up to the scaling limit of the specific parallel strategy (e.g., maximum number of filters). More specifically, we scale the tests from $16$ to $1024$ GPUs for data and hybrid parallelism, from $4$ to $64$ GPUs for filter/channel parallelism, and up to $4$ GPUs for pipeline parallelism. The y-axis shows the iteration time for each case. The iteration time is calculated as an average of $100$ iterations excluding the first iteration which normally involves initialization tasks. To get a more detailed analysis, we decompose the execution time into computation and communication. The oracle prediction is shown in blue as stacked bars, i.e., computation+communication, and the measured empirical results are shown in orange. In this figure, we report the best communication times obtained during our experiments, as this represents the peak performance the hardware can deliver and leave aside occasional delays due to external factors such as network congestion coming from other apps, system noise and, overheads due to correctable errors, among others. A detailed analysis on network congestion is given on Section~\ref{sec:result_communication}. The labels above each column show the \emph{projection accuracy} in percentage, i.e., 1 - ratio of the absolute value of the difference with respect to the total measured time. Similarly, Figure~\ref{fig:accuracy_cosmoflow} shows the accuracy for CosmoFlow in the case of Data+Spatial parallelism. \leo{Note that the reason CosmoFlow is only run on the Data+Spatial hybrid configurations is because the sample size is so large that it cannot run with any other parallel strategy.}

The accuracy of ParaDL predictions for the different parallel strategies are 96.10\% for data parallelism, 85.56\% for Filter, 73.67\% for Channel, 91.43\% for Data+Filter, 83.46\% for Data+Spatial and 90.22\% for pipeline across all CNN models. In general, this represents an overall accuracy of \aveacu for ParaDL, across all parallelism strategies and CNN models, and up to \maxacu for \bestpar  on VGG16. CosmoFlow shows an accuracy of 74.14\% on average.

It is important to note that the overall average accuracy drops significantly due to some few outliers in which the communication time measured is substantially higher than the prediction from ParaDL. For instance Data+Spatial for ResNet-152 with $512$ GPUs \leo{shows an accuracy of 62\% due to network congestions}. Interestingly, the same configuration with $1024$ GPUs shows a much higher accuracy (i.e., 83\% for Data+Spatial ResNet-152) including the communication part, demonstrating that the ParaDL oracle is highly accurate, even at large scale (i.e., $1024$ GPUs). 
Section~\ref{sec:result_communication} includes a detailed analysis of the network congestion leading to the few outliers 
%(i.e., Data+Filter ResNet-50 $512$ GPUs), 
where the machine was oversubscribed.
%and suffered from significant network congestion.
Note that the communication time of ParaDL for Data+Filter shown in Figure~\ref{fig:comparision} is calculated with a contention penalty coefficient of $2\times$, e.g., contention caused by two disjoint Allreduces that share the same InfiniBand link for inter-node communication. The high accuracy reported show that our analytical model fits well to the real performance.

\subsection{Parallelism Limitations and Bottlenecks}
\label{sec:result_limit}
\begin{figure*}
	\begin{minipage}{0.3\linewidth}
		\begin{figure}[H]
			\centering
			\includegraphics[width=0.9\linewidth]{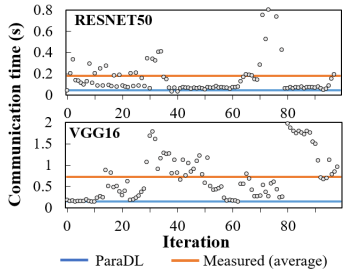}
			\caption{\small{Network congestion of ResNet-50, 512 GPUs, data parallelism (upper) and VGG16, 64 GPUs, filter parallelism (lower).}}
			\label{fig:congestion}
		\end{figure}
	\end{minipage}
	\hspace{0.5cm}
	\begin{minipage}{0.3\linewidth}
		\begin{figure}[H]
			\includegraphics[width=0.9\linewidth]{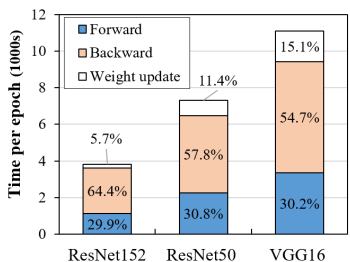}
			\caption{\small{Computation time per epoch with PyTorch. Weight update is not trivial in large models and dataset.}}
			\label{fig:weightupdate_big}
		\end{figure}
	\end{minipage}
\hspace{0.5cm}
	\begin{minipage}{0.32\linewidth}
	\begin{figure}[H]
		\includegraphics[width=0.85\linewidth]{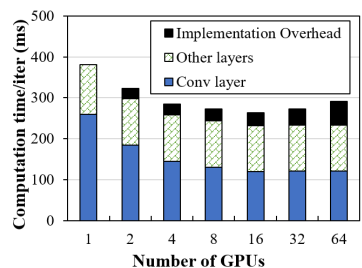}
		\caption{\small{ Computation breakdown of filter parallelism, ResNet-50. Implementation of convolution layers does not scale well.}}
		\label{fig:filter_resnet_comp}
	\end{figure}
\end{minipage}
	\vspace{-0.3cm}
\end{figure*}

In this section we use a combination of observations from ParaDL projections and empirical results to highlight some important points: (i) inherent to the parallel strategies themselves (limitations), and (ii) those caused by other components such as the framework implementation or system architecture (bottlenecks).
This helps users in avoiding these limitations, and framework programmers in prioritizing their efforts for improvements.
We group these limitations and bottlenecks into four categories.

\subsubsection{\textbf{Communication}}
\label{sec:result_communication}
It is well-known in literature that parallel training introduces communication overhead.
Those overheads have different forms and patterns for different parallel strategies. There is the gradient exchange at the end of each iteration in data and spatial parallelism (GE-Allreduce). There can also be extra communication in the forward and backward passes of other parallel strategies: the layer-wise collective communication in filter/channel (FB-Allgather and FB-Allreduce), layer-to-layer communication in pipeline (FB-layer), and the halo exchange in spatial (FB-Halo).

\textit{Gradient Exchange:} Similar to data parallelism, spatial requires a gradient exchange to aggregate the weights. This collective communication, i.e., Allreduce, has significant impact on performance, and can become a limitation.
Another point worth mentioning is the hierarchal implementation of Allreduce in hybrid strategies such as Data+Filter (\textit{df}) and Data+Spatial (\textit{ds}).
These two types of Allreduce (data and hybrid parallelism) are different as they involve different parallelization techniques. 
In \textit{ds}, we perform a reduce inside each node to leader GPUs first, then perform a global Allreduce between the leaders. 
However such implementation leads to a higher overhead, e.g., time for Allreduce  is more than $2\times$ as those of data. Alternative ways to address this issue are to use multiple leaders instead of only one~\cite{nguyen2018hierarchical} or to use segmented allreduces, i.e., smaller, concurrent allreduces among disjoint sets of GPUs~\cite{Naoya:sc19}. We use the former strategy for \emph{ds} and the latter for \emph{df}. These methods are not trivial to implement and they require significant engineering effort. ParaDL has proven accurate at modeling those communications and can be used to choose an implementation strategy based on ParaDL's projected communication times.  

\textit{Layer-wise collective communication:} unlike data parallelism, filter and channel parallelism require multiple collective communication rounds at each layer. The communication time depends on the activation size $\times$ batch size, i.e., $\mathcal{O}(B\sum_{l=1}^{G}|y_l|)$, as well as the depth of DL model, i.e., $\sum_{l=1}^{G}(p-1)\alpha$ as reported in our analytic model. In our experiments with ImageNet, even though the total activation sizes are smaller than the number of weights, yet with a batch size of $\ge 32$ samples, the communication time of filter/channel is larger than that of data parallelism (See Figure~\ref{fig:comparision}). Note that because this communication overhead is attributed to the forward and backward phases, Allreduce optimization techniques 
such as sparsification
are no-longer valid. Instead, a hybrid which combines filter and channel (plus data) parallelism may help to mitigate this limitation by reducing the number of communication calls~\cite{Shoeybi2019MegatronLMTM} or using a smaller segmented Reduce-Scatter~\cite{Naoya:sc19}.

\textit{Peer to Peer communication}: The halo exchange in spatial parallelism and the activation passing between composite layers in pipeline are performed in a P2P fashion, which are  expected to have small communication times.
However, ParaDL shows that the communication time of FB-Halo is non-trivial and this was confirmed by the empirical results. For example, in ResNet-50, 128 GPUs, the time of FB-Halo is approximately $60\%$ of the gradient exchange Allreduce, which is substantially higher than initially expected. 
This bottleneck appears because the framework uses the MPI library instead of NCCL (NCCL allows GPUs to communicate directly instead of via CPU, i.e., GPUDirect). We plugged different network parameters in ParaDL (See Section~\ref{sec:paradl:perf}) for MPI and NCCL and we confirmed the difference, both theoretically and empirically.

\textit{Network Congestion:} %and network congestion
In our empirical experiments, we try to avoid the issue of congestion as much as possible by running several times for each data point. However, as shown in Figure~\ref{fig:comparision}, we still observe network congestion when approaching 1K GPUs. 
We did a detailed analysis for several of the runs. In Figure~\ref{fig:congestion} we show the time for Allreduce communication for data parallelism of ResNet-50 with $512$ GPUs and an Allgather communication for filter parallelism of VGG16 with $64$ GPUs.
We noticed that most data points align well with the expected theoretical bandwidth predicted by our analytical model (blue line), yet network congestion caused by other jobs in the system can lead to some outliers that push the average communication time up to four times higher than expected. This overhead can not be avoided especially for large-scale training, e.g., 100s-1000s GPU, in a shared HPC system.
It could, however, be mitigated at the system level by switching to a full-bisection bandwidth rather than having 1:3 over-subscription.

\subsubsection{\textbf{Memory Capacity}}
\label{sec:result_memory}
We highlight specific cases when memory requirements become an issue in different strategies.

\textit{Redundancy in Memory}: Different memory redundancies could emerge in different parallel strategies. In the spatial and channel/filter strategies the activations (i.e. input/output channels) are divided among nodes, however this does not reduce the memory requirements of holding the weight tensors since their weights are not divided among PEs (as in our analytical model). This becomes an issue for larger models.
One alternative proposed in ZeRO~\cite{Rajbhandari2019ZeROMO} is to split the weights as well as the activations. However, this comes at the cost of extra communication of 50\% since two Allgathers of the weights are needed in the forward and backward passes. In pipeline, the memory required for a single layer could be prohibitive. For example, ComsoFlow's first Conv layer generates more than 10GB of activation tensor when the input size is $4\times 512^3$. Accordingly, for those kind of models the pipeline strategy would be unfeasible and one has to resort to other parallel strategies (e.g., use Data+Spatial for CosmoFlow as shown in Figure~\ref{fig:ds_CosmoFlow}). Additionally, since samples are fed in a pipelined fashion, the memory required is proportional to the number of stages, and would hence become a bottleneck in deep pipelines, unless we apply gradient checkpointing at the boundary of the partition~\cite{huang2018gpipe, pipedream}, which comes with the overhead of recomputing the activations within each partition.

\wahib{\textit{Memory Manager}: DL frameworks typically include a memory manager to reduce the overhead of frequent malloc and free. Since GPUs typically operate in asynchronous execution model, there are many CUDA kernels being launched at any given time in training. We observed a disparity between ParaDL and measured performance that could be attributed to stalling kernels. Upon inspection, we observed kernels that were launched asynchronously often stall when requesting memory (in, Chainer as well as PyTorch). The launched asynchronously launched kernels waiting for memory to be available leads to heavy fluctuations in performance. Guided by ParaDL to identify the fine-grained location of the performance gap, we confirmed, for instance, that the implementation of Data+Spatial parallelism of VGG16 ($64$ GPUs) could avoid out-of-memory issues at the cost of a performance degradation of 1.5$\times$.}

\subsubsection{\textbf{Computation}}
\label{sec:result_computation}
We highlight the following limitations.

\textit{Weight update}: most compute time in training typically goes to the forward and backward pass. However, we observed with our analytical model that for larger models the weight update starts to become a significant portion of the compute time. For instance, we measured weight update to take up to 15\% for the VGG16
(shown in Figure~\ref{fig:weightupdate_big}). Larger DL models, using ADAM optimizer in specific, may need a higher computation time for weight. 
Especially, large Transformer
based models report up to 45\% time on weight update and more than 60\% extra memory requirements since ADAM requires four variables per weight~\cite{wahib2020scaling}. One alternative to address this is to shard the weight update among GPUs across iterations, and Allgather the weights before forward/backward passes~\cite{xu2020automatic}.

\textit{Workload Balancing:}
Pipeline can outperform data parallelism with less communication by using P2P rather than a collective communication. However, it is crucial that all stages in the pipeline take roughly the same amount of time, since the training time of a pipeline is limited by the slowest stage. Indeed, there may be cases where no simple partitioning across the GPUs achieves perfect load balance (e.g. networks with non-linear connections). To further improve load balancing, a straight forward approach is to use data parallelism inside a stage, i.e., hybrid of pipeline plus data~\cite{pipedream}.

\textit{Computation Redundancy:} This section discusses limits that could arise from computational redundancy that is introduced for different parallel strategies.
Using ParaDL we found out that there was a gap between analytical result and the measured time in filter/channel. Looking in detail, we saw that this was an implementation issue in the framework including two factors i) the convolution layer does not always scale as expected and ii) the computation overhead, such as split/concat, is non-trivial.
These non-trivial implementation overheads are shown in Figure~\ref{fig:filter_resnet_comp}.

\subsubsection{\textbf{Scaling limitation}}
\label{sec:result_scaling}
When scaling, there is a limit on the number of GPUs for each of the model parallel strategies (last column of Table~\ref{table:sum_time}). For example, $p$ can not exceed the minimum number of filters of a layer in the model, i.e., $64$ in the case of VGG16 and ResNet-50 with filter parallelism. Hybrid approaches have a better scaling than those of pure model parallelisms. For example, as shown in Figure~\ref{fig:ds_CosmoFlow}, using Spatial+Data hybrid is an effective scalable alternative (despite communication inefficiencies), since it scales both in performance and GPU count (i.e. one could simply expand the data parallel pool as much as new nodes are added). Indeed, the curve shows a perfect scaling (note the logarithmic y-axis). 

\subsection{Other Observations}
This section discusses a few more points related to CNN training. 

\subsubsection{The Rise of Hybrid Parallelism}
As mentioned, each of four basic parallelism strategies has its own limitations. Using the hybrid strategies (Data+Model) helps to break/mitigate those limitations,  
e.g., memory issue of data parallelism and communication and scaling limitation of model parallelism (Section~\ref{sec:result_limit}).
As more datasets from the HPC field start to be trained by DL frameworks, this type of hybrid parallel strategies will become increasingly relevant because data parallelism will simply be not enough, as shown in the case of CosmoFlow and its good scaling with \emph{ds} in Figure~\ref{fig:ds_CosmoFlow}.
In addition, hybrid strategies may drive to a better solution in terms of performance.
In accordance with other recent reports~\cite{Naoya:sc19},
there are cases where data+filter (\textit{df}) hybrid can outperform data parallelism at large scale, as we also observed in some of our experiments (we also noticed scenarios where pipeline outperforms data parallelism).

\subsubsection{Distributed Inference}
Inference at large scale is becoming increasingly demanded, given that for large models the inference would also be distributed~\cite{brown2020language}. In smaller models, when latency of inference matters in an application, the inference could also be distributed (e.g., real-time prediction of Tokamak disruptions in magnetically-confined thermonuclear plasma experiments~\cite{dong2020fully}). Some of the limitations and bottlenecks of distributed training discussed previously also appear in distributed inference (See lines marked with \textbf{Y} in column \textbf{I} of Table~\ref{table:limit_summary}).

\begin{table}[bt]
	\caption{\small{Summary of detected limitations (\textbf{L}) and bottlenecks (\textbf{B}) with the related training phases ($\checkmark$) and components ($\sbullet[1.5]$). Those limitations/bottlenecks may appear (Y/N) in distributed inference (\textbf{I}). Related parallel strategies ($\times$): \textbf{d}-data, \textbf{s}-spatial, \textbf{p}-pipeline, \textbf{f/c}-filter and channel, \textbf{df}-hybrid Data+Filter, \textbf{ds}-hybrid Data+Spatial. \textbf{FR}-Framework, \textbf{SY}-System.} Training phases: \textbf{IO}-I/O and pre-processing, \textbf{FB}-a forward and backward propagation, \textbf{GE}-the gradient exchange (if needed) and \textbf{WU}-updating the weights.} 
	\label{table:limit_summary}
	\centering
	\scriptsize
	\setlength\tabcolsep{3pt} % default value: 6pt
	\resizebox{0.5\textwidth}{!}{%
		\begin{tabular}{lccccccccccccccl}
			\toprule
			\multirow{2}{*}{\textbf{\begin{tabular}[c]{@{}c@{}}Category\end{tabular}}} & \multirow{2}{*}{\textbf{\begin{tabular}[c]{@{}c@{}}L/B\end{tabular}}} &
			\multicolumn{6}{c}{\textbf{\begin{tabular}[c]{@{}c@{}}\underline{P}ara. \underline{S}trategies\end{tabular}}} &
			\multirow{2}{*}{\textbf{\begin{tabular}[c]{@{}c@{}}FR\end{tabular}}} &
			\multirow{2}{*}{\textbf{\begin{tabular}[c]{@{}c@{}}SY\end{tabular}}} &
			\multicolumn{4}{c}{\textbf{Training Phases}}
			&\multirow{2}{*}{\textbf{I}}
			&\multirow{2}{*}{\textbf{Remarks}} 
			\\ \cmidrule(lr){3-8} \cmidrule(lr){11-14}
			& & \textbf{d} & \textbf{s} & \textbf{p}  & \textbf{f/c} &\textbf{df}  & \textbf{ds}  & & &\textbf{IO} & \textbf{FB} & \textbf{GE} & \textbf{WU}  \\ \midrule \rC
			
			%& B & $\times$ & $\times$ & - & - & - & $\times$ & $\sbullet[1.5]$& $\sbullet[1.5]$ & - & $\checkmark$ &  - & $\checkmark$ & Y &Not enough concurrency\\ 
			
			\WC & \WC L & \WC $\times$ & \WC $\times$ & \WC - & \WC - & \WC $\times$ & \WC $\times$ & \WC$\scirc[1.5]$ & \WC $\scirc[1.5]$ & \WC - &\WC  - & \WC $\checkmark$ & \WC - &\WC N & \WC Gradient-exchange \\ 
			
			\WC& \WC L & \WC - & \WC- & \WC- & \WC$\times$ & \WC$\times$ & \WC- & \WC $\scirc[1.5]$ & \WC$\scirc[1.5]$ & \WC- & \WC$\checkmark$ &  \WC- & \WC- &\WC Y & \WC Layer-wise comm.\\
			
			\WC & \WC B & \WC - & \WC $\times$ & \WC $\times$ & \WC - & \WC - & \WC $\times$ & \WC$\sbullet[1.5]$ & \WC $\scirc[1.5]$ & \WC - &\WC  $\checkmark$ & \WC $\checkmark$ & \WC - &\WC Y & \WC P2P communication\\
			
			\WC\multirow{-4}{*}{\begin{tabular}[c]{@{}l@{}}Commu-\\nication\end{tabular} }& \WC B & \WC $\times$ & \WC $\times$ & \WC $\times$ & \WC $\times$ & \WC $\times$ & \WC $\times$ & \WC$\scirc[1.5]$ & \WC $\sbullet[1.5]$ & \WC - &\WC  $\checkmark$ & \WC $\checkmark$ & \WC - &\WC Y & \WC Network Congestion\\ 
						
			& B & $\times$ & $\times$ & $\times$ & $\times$ & $\times$ & $\times$ & $\scirc[1.5]$& $\sbullet[1.5]$&  $\checkmark$  &  $\checkmark$ &  $\checkmark$ &   $\checkmark$ & Y & Memory Redundancy\\
			%& B & $\times$ & $\times$ & $\times$ & $\times$ & $\times$ & $\times$ & $\sbullet[1.5]$ & $\scirc[1.5]$&  $\checkmark$  &  $\checkmark$ &  $\checkmark$ & $\checkmark$ &Y & Fragmentation\\  \rC
			\multirow{-2}{*}{\begin{tabular}[c]{@{}l@{}} Memory \\Capacity
			\end{tabular}} & B & $\times$ & $\times$ & $\times$ & $\times$ & $\times$ & $\times$ & $\sbullet[1.5]$ & $\scirc[1.5]$&  $\checkmark$  &  $\checkmark$ &  $\checkmark$ & $\checkmark$ &Y & Memory Stalling\\  \rC
			
			 & L & $\times$ & $\times$ & $\times$ & $\times$ & $\times$ & $\times$ & $\scirc[1.5]$& $\scirc[1.5]$ & - & - &  - & $\checkmark$ & N &Weight Update\\
			 
			\WC & \WC L & \WC- & \WC- & \WC$\times$ & \WC- & \WC- & \WC- & \WC$\scirc[1.5]$& \WC$\scirc[1.5]$ & \WC- & \WC$\checkmark$ & \WC- & \WC$\checkmark$ & \WC Y & \WC Workload Balancing\\
			
			\WC\multirow{-3}{*}{\begin{tabular}[c]{@{}l@{}}Comput-\\-ation\end{tabular} } & \WC B & \WC- & \WC- & \WC- & \WC$\times$ & \WC$\times$ & \WC- & \WC$\sbullet[1.5]$& \WC$\scirc[1.5]$ & \WC- & \WC$\checkmark$ &  \WC- & \WC$\checkmark$ &\WC Y & \WC Comp. Redundancy\\ 
			
			\begin{tabular}[c]{@{}l@{}} Scaling
			\end{tabular} & L & $\times$ & $\times$ & $\times$ & $\times$ & $\times$ & $\times$ &  $\scirc[1.5]$ & $\scirc[1.5]$ &  $\checkmark$  &  $\checkmark$ &  $\checkmark$ &   $\checkmark$ &Y & Number of PEs \\ 
			
			\bottomrule
		\end{tabular}
	}
	\vspace{-0.3cm}
\end{table}

\section{Conclusion}
\label{sec:conclusion}
We propose an analytical model for characterizing and identifying the best technique of different parallel strategies for CNN distributed training.
We run a wide range of experiments with different models, different parallel strategies and different datasets for up to 1,000s of GPUs and compare with our analytical model. The results demonstrate the accuracy of ParaDL, as high as \maxacu, and \aveacu on average accuracy across all parallel strategies on multiple CNN models and datasets on up to 1K GPUs. 

The analytical model helped us uncover limitations and bottlenecks of parallel training of CNNs (summary in Table~\ref{table:limit_summary}). We analytically identify different bottlenecks that can appear in different parallel strategies due to communication patterns that compensate for different ways to split the tensors, and confirm those predictions empirically. Finally, we identify memory and computational pressure that arises from different redundancies in different parallel strategies. 

\section*{Acknowledgment}
The project that gave rise to these results received the support of a fellowship from the "la Caixa" Foundation (ID \textbf{ 100010434}). The fellowship code is \textbf{LCF/BQ/DI17/11620059}.

This project has received funding from the European Union's Horizon 2020 research and innovation programme under the Marie Skłodowska-Curie grant agreement No. \textbf{713673}.

The Eurolab4HPC project has received funding from the European Union Horizon 2020 Framework Programme (H2020-EU.1.2.2. - FET Proactive) under grant agreement number \textbf{800962}

This work was supported by JST, ACT-X Grant Number JPMJAX190C, Japan; by JST, PRESTO Grant Number JPMJPR20MA, Japan.

\bibliographystyle{ACM-Reference-Format}
\bibliography{refer}

\newpage

\appendix
 \section{Appendix}
 \subsection{Performance and Memory Analysis}
The training time of one epoch in the sequential implementation (serial) of a CNN includes only the time for computation:
\begin{equation}
\label{equal:orig_time}
\scriptsize
\begin{split}
T_{serial} = \sum_{1}^{I}B\sum_{l= 1}^{G}\Big(FW_l + BW_l\Big) + \sum_{1}^{I}\sum_{l= 1}^{G}\Big(WU_l\Big) \\
= D\sum_{l= 1}^{G}\Big(FW_l + BW_l\Big) + \frac{D}{B}\sum_{l= 1}^{G}(WU_l)
\end{split}
\end{equation} 
Considering the memory footprint:
\begin{equation}
\label{equal:orig_mem}
\scriptsize
\begin{split}
M_{serial} = 2\delta\sum_{l=1}^{G}\Big(B(|x_l| + |y_l|) + |w_l| \Big)
\end{split}\end{equation}	
In the following, we estimate the total training time and maximum memory per PE for the mentioned basic parallelism strategies and one hybrid strategy.

\subsubsection{Data parallelism}
In this strategy, the training time includes both computation and communication time.
Each PE processes a micro batch size $B' = \frac{B}{p}$ in this case. The time for computing at layer $l$ in one iteration for forward and backward phase is $\frac{1}{p}$ of the single-process.
Thus the total computation time in one epoch becomes:
\begin{equation}
\label{equal:data_comp_3}
\scriptsize
\begin{split}
T_{data,comp} = \sum_{1}^{I}\sum_{l = 1}^{G}\Big(\frac{B}{p}(FW_l + BW_l) + WU_l \Big)  \\
= \frac{D}{p}\sum_{l = 1}^{G}(FW_l + BW_l) + \frac{D}{B}\sum_{l = 1}^{G}(WU_l)
\end{split}
\end{equation}

Because PEs have to share their gradients at the end of each iteration, the time for communication is $\frac{D}{B}T_{ar}(p,\sum_{l=1}^{G}|w_l|)$.
i.e., an Allreduce operation with a ring-based algorithm, the time for communication is:
\begin{equation}
\label{equal:data_comm}
\scriptsize
\begin{split}
T_{data,comm} % =\sum_{1}^{I}2(p-1)(\alpha + \frac{\sum_{l=1}^{G}|w_l|}{p}\beta)\\
= 2\frac{D}{B}(p-1)\Big(\alpha + \frac{\sum_{l=1}^{G}|w_l|}{p}\delta\beta\Big)
\end{split}
\end{equation}
Clearly, data parallelism has the benefit of reduction in computation time by $\frac{1}{p}$ at the price of communication time. 

Considering the memory footprint, in data parallelism we duplicate the entire model on $p$ different PEs. Each PE processes a partition of the dataset in a microbatch of $B'=\frac{B}{p}$ samples.
A layer $l$ needs memory to store its input $B'|x_l|$, activation $B'|y_l|$, weights $|w_l|$, \nguyen{bias $|bi_l|$}, the gradients $B'|\frac{dL}{dx_l}|$, $B'|\frac{dL}{dy_l}|$, and $|\frac{dL}{dw_l}|$.
%Some models with a huge number of parameters may require significant memory for "bias" such as in a fully-connected layer. We consider to not formally show such memory requirement to make our analysis easy to follow. 
Overall, if each item of the input, activation, weight and gradients are stored in $\delta$ bytes, 
%the total memory of the sequential implementation is $\sum_{l=1}^{G}\delta(B'(|x_l|+|y_l|)+ |w_l| + B'(|\frac{dL}{dx_l}|+|\frac{dL}{dy_l}|) + |\frac{dL}{dw_l}|) $.Thus 
the maximum required memory at one PE is: %$2\delta\sum_{l=1}^{G}(\frac{B}{p}(|x_l| + |y_l|) + |w_l| )$. %:
\begin{equation}
\label{equal:data_mem_2}
\scriptsize
\begin{split}
M_{data} = \sum_{l=1}^{G}\delta(B'(|x_l|+|y_l|)+ |w_l| + \nguyen{|bi_l|} +  B'(|\frac{dL}{dx_l}|+|\frac{dL}{dy_l}|) + |\frac{dL}{dw_l}|)
\\ 
=  \delta\sum_{l=1}^{G}\Big(2\frac{B}{p}(|x_l| + |y_l|) + 2|w_l| + \nguyen{|bi_l|}\Big)
\end{split}
\end{equation} 
\subsubsection{Spatial parallelism}
As mentioned in the previous section, the spatial dimensions of $x$, $y$, $\frac{dL}{dx}$ and $\frac{dL}{dy}$ are split among $p$ PEs so that the memory at one PE is: %shown in Equation~\ref{equal:spatial_mem}.
%:
%NguyenTT: Comment for reused later
\begin{equation}
\label{equal:spatial_mem}
\scriptsize
\begin{split}
M_{spatial} = \delta\sum_{l=1}^{G}\Big(2B\frac{(|x_l| + |y_l|)}{p} + 2|w_l| + \nguyen{|bi_l|}\Big)
\end{split}
\end{equation}
Because each PE performs a computation with the size of the spatial dimensions as a fraction $\frac{1}{pw}$, $\frac{1}{ph}$, and $\frac{1}{pd}$ of the sequential implementation. This reduces the computation time of forward and backward phase of a layer by $p = pw\times ph\times pd$ times %(Equation~\ref{equal:spatial_comp}).
%NguyenTT: Comment for reused later
Thus, the computation time is:
\begin{equation}
\label{equal:spatial_comp}
\scriptsize
\begin{split}
T_{spatial,comp} = \sum_{1}^{I}B\sum_{l = 1}^{G}(\frac{FW_l}{p} + \frac{BW_l}{p}) + \sum_{1}^{I}\sum_{l= 1}^{G}\Big(WU_l\Big)\\
= \frac{D}{p}\sum_{l = 1}^{G}(FW_l + BW_l) + \frac{D}{B}\sum_{l = 1}^{G}(WU_l)
\end{split}
\end{equation}
The communication time includes the time to perform the Allreduce operation to share the weight gradients (similar to data parallelism) and the time to perform the halo exchange of each layer.
For a layer $l$, a PE needs to send/receive the halo regions with the logically-neighboring PE(s). 
Thus the total time for halo exchange is %$2\frac{D}{B}\sum_{l=1}^{G}(T_{p2p}(\text{halo}(|x_l|)) + T_{p2p}(\text{halo}(|\frac{dL}{dy_l}|)))$.
\begin{equation}
\label{equal:spatial_comm_halo}
\scriptsize
\begin{split}
T_{spatial,halo} = 2\frac{D}{B}\sum_{l=1}^{G}(T_{p2p}(B(\text{halo}(|x_l|))) + T_{p2p}(B(\text{halo}(|\frac{dL}{dy_l}|))))\\
= 2\frac{D}{B}\sum_{l=1}^{G}(2\alpha +B\delta\beta(\text{halo}(|x_l|) + \text{halo}(|\frac{dL}{dy_l}|)))
\end{split}
\end{equation}

In which $\text{halo}()$ presents the size of data exchanged per batch. The exchanged data size depends on how each spatial dimension is split.

\subsubsection{Layer parallelism}
In this strategy, a DNN model is split into $p$ composite layers (or group). 
Let $g_i$ denote the group assigned to PE $i$. 
That is, each PE $i$ keeps $G_i$ layers of the model given that $\sum_{i=1}^{p}G_i = G$.
Let $FW_{G_i}$, $BW_{G_i}$, and $WU_{G_i}$ denote the time for performing the forward, backward, and weight update computation of group $i$, i.e., $FW_{G_i} = \sum_{l\in g_i}(FW_l)$, $BW_{G_i} = \sum_{l\in g_i}(BW_l)$, and $WU_{G_i} = \sum_{l\in g_i}(WU_l)$.

\textit{Pure implementation} processes a batch of $B$ samples at the first node and then sequentially pass the intermediate activation (gradients) through all $p$ nodes in each iteration. Hence, the time for computation is: 
\begin{equation}
\label{equal:layer_comp}
\scriptsize
\begin{split}
T_{layer,comp} = \sum_{1}^{I}(B\sum_{i = 1}^{p}(FW_{G_i} + BW_{G_i}) + \sum_{i = 1}^{p}(WU_{G_i})) 
\\
=  D\sum_{l=1}^{G}(FW_l + BW_l)+ \frac{D}{B}\sum_{l = 1}^{G}(WU_l)
\end{split}
\end{equation}
This approach does not reduce the computation time but it is helpful if the memory footprint at one node is limited.
In practice, a pipeline implementation is used to reduce the computation time.

In a \textit{pipeline implementation}, the mini-batch is divided into $S$ segments of size $\frac{B}{S}$.  
In one stage, the computation of a layer group (or PE) $g_i$ on a data segment $s$ is performed simultaneously with the computation of layer group $g_{i+1}$ on the data segment $s-1$, and so on. 
Thus, the time for each stage can be approximated by the maximum computation time of layer groups, i.e., $\max_{i = 1}^{p}(FW_{G_i})$ or $\max_{i = 1}^{p}(BW_{G_i})$. In general, a pipeline implementation of $p$ PEs with $S$ data segments requires $(p + S-1)$ stages per iteration that leads to the total computation time of one epoch as: % in Equation~\ref{equal:layer_comp}.

\begin{equation}
\label{equal:pipe_comp}
\scriptsize
\begin{split}
T_{pipe,comp} \approx %\sum_{1}^{I}\frac{B}{S} (p + S-1)\Big(\max_{i = 1}^{p}(FW_{G_i}) +\max_{i = 1}^{p}(BW_{G_i})\Big)\\
%=  
\frac{D(p + S-1)}{S}(\max_{i = 1}^{p}(FW_{G_i}) +\max_{i = 1}^{p}(BW_{G_i} +\max_{i = 1}^{p}(WU_{G_i})) 
\end{split}\end{equation}

%%NguyenTT: Comment for reused later
%In case that the computation time on all PEs is evenly balanced, e.g., $\frac{1}{p}$ the computation of all the layers per PE, as mentioned in~\cite{huang2018gpipe}, Equation~\ref{equal:pipe_comp} can be simplified as:
%\begin{equation}
%\label{equal:pipe_comp_2}
%\scriptsize
%\begin{split}
%T_{pipe,comp} \approx \frac{(p + S-1)D}{S}\sum_{l = 1}^{G}\frac{1}{p}\Big(FW_l + BW_l\Big) 
%\end{split}\end{equation}

Considering the communication in this strategy, each PE $i$ has to pass forward/backward the output/input's gradients to the next/previous PE in a peer-to-peer communication scheme which costs $T_{p2p}(B|y_{G_i}|)$ and $T_{p2p}(B|\frac{dL}{dx_{G_i}}|)$, where $y_{G_i}$ and $x_{G_i}$ denote the output of the last layer and input of the first layer of a group layer $g_i$, respectively.
%$\alpha + B|y_{G_i}|\delta\beta$ and $\alpha + B|\frac{dL}{dx_{G_i}}|\delta\beta$, where $y_{G_i}$ and $x_{G_i}$ denote the output of the last layer and input of the first layer of a group layer $g_i$, respectively.
In the pipeline fashion, the communication time of each stage can be approximated by $\max_{i =1}^{p-1}{T_{p2p}(B|y_{G_i}|)}$ and $\max_{i=2}^{p}{T_{p2p}(B|\frac{dL}{dx_{G_i}}|)}$. 
%%NguyenTT: Comment for reused later
%The total communication time of one iteration thus becomes:
%\begin{equation}
%\scriptsize
%\begin{split}
%(p + S - 2)\big( \max_{i=1}^{p-1}(\alpha + \frac{B}{S}|y_{G_i}|\delta\beta) + \max_{i=2}^{p}(\alpha + \frac{B}{S}|\frac{dL}{dx_{G_i}}|\delta\beta)\big)
%\end{split}\end{equation}
In the case of $|x_l| = |y_{l-1}|$, the total time for communication in one epoch ($I = \frac{D}{B}$ iterations) is summarized in: % the Equation~\ref{equal:layer_comm}. 
%%NguyenTT: Comment for reused later
\begin{equation}
\label{equal:pipe_comm}
\scriptsize
\begin{split}
T_{pipe,comm} \approx 2\frac{D(p + S - 2)}{B}\Big( \max_{i =1}^{p-1}\big(\alpha + \frac{B}{S}|y_{G_i}|\delta\beta\big)\Big)
\end{split}\end{equation}

For the memory footprint, because each PE $i$ stores a different set of layers, the maximum required memory in one PE is: %as in Equation \ref{equal:layer_mem}.
%$2\delta \max_{i=1}^{p}(2\sum_{l=1}^{G_i}(B(|x_l| + |y_l|) + |w_l| ))$. %:
%%NguyenTT: Comment for reused later
\begin{equation}
\label{equal:pipe_mem}
\scriptsize
\begin{split}
M_{pipe} = \delta \max_{i=1}^{p}\Big(\sum_{l=1}^{G_i}\big(2B(|x_l| + |y_l|) + 2|w_l| + \nguyen{|bi_l|}\big)\Big)
\end{split}
\end{equation}

\subsubsection{Filter parallelism}
In this strategy,
%we define the computation time as follows:
%\begin{equation}
%\label{equal:filter_comp}
%\small
%\begin{split}
%T_{filter,comp} = \frac{D}{p}\sum_{l = 1}^{G}\Big(FW_l + BW_l\Big)
%\end{split}
%\end{equation}
the computation time is reduced $p$ times,  %(Equation \ref{equal:filter_comp})
yet the time for communication at a layer $l$ becomes more complex, since it includes
(1) an Allgather at the forward phase (except layer $G$)\footnote{Each process $i$ transfers $|(y_l)_i[*,p,*]| = \frac{|y_l|}{p}$ values for one sample in layer $l$, and a total of $B\frac{|y_l|}{p}$ values for the entire batch.} that costs $T_{ag}(p,\frac{B|y_l|}{p})$, 
%\begin{equation}
%\small
%%\sum_{l=1}^{G-1}
%(p-1)(\alpha + \frac{B|y_l|}{p}\delta\beta) %= (p-1)(\alpha + \frac{\sum_{l=1}^{G-1}|y_l|}{p}\delta\beta)
%\end{equation}
and (2) an Allreduce at the backward phase (except layer 1) that costs $T_{ar}(p,B|\frac{dL}{dx_l}|) = T_{ar}(p,B|x_l|)$.
%\begin{equation}
%\small
%%\sum_{l=1}^{G}
%2(p-1)(\alpha + \frac{B|\frac{dL}{dx_l}|}{p}\delta\beta) = 2(p-1)(\alpha + \frac{B|x_l|}{p}\delta\beta)
%\end{equation}
%%NguyenTT: Comment for reused later
%Thus, the total time for communication in one iteration is: 
%\begin{equation}
%\scriptsize
%\begin{split}
%\sum_{l=1}^{G-1}(p-1)(\alpha + \frac{B|y_l|}{p}\delta\beta) + \sum_{l=2}^{G}2(p-1)(\alpha + \frac{B|x_l|}{p}\delta\beta)
%\end{split}
%\end{equation}
In the case of $|x_l| = |y_{l-1}|$, the total time for communication in $I = \frac{D}{B}$ iterations is: % shown in Equation~\ref{equal:filter_comm}. 
%%NguyenTT: Comment for reused later:
\begin{equation}
\label{equal:filter_comm1}
\scriptsize
\begin{split}
T_{filter,comm} = 3\frac{D}{B}(p-1)\sum_{l=1}^{G-1}(\alpha + \frac{B|y_l|}{p}\delta\beta)
%\\ = 3\frac{D}{B}(p-1)\Big((G-1)\alpha + \frac{B\sum_{l=1}^{G-1}|y_l|}{p}\delta\beta\Big)
\end{split}
\end{equation}

%The total training time becomes:
%\begin{equation}
%\small
%\begin{split}
%\label{equal:filter_time}
%T_{filter} = \frac{D}{p}\sum_{l=1}^{G}(FW_l + BW_l) 
%+ 3\frac{D}{B}\sum_{l=1}^{G-1} \Big((p-1)(\alpha + \frac{B|y_l|}{p}\delta\beta)\Big)
%\end{split}
%\end{equation}

In this strategy, each PE keeps only $\frac{1}{p}$ the filters (weight) of each layer.
%A given node $i$ also computes the activation of layer $k$ partially, i.e., $Y_{i,k} \gets WT_{i,k} * X_{i,k}$ where $|WT_{i,k}| = \frac{|WT_k|}{p}$ in the forward phase.
However, PE $i$ needs to communicate with other PEs to share its local partial activations, hence requiring memory to store the entire activation $|y_k|$. 
The required memory at each PE is: % presented in Equation~\ref{equal:filter_mem}. %:
%%NguyenTT: Comment for reused later
\begin{equation}
\scriptsize
\label{equal:filter_mem}
\begin{split}
M_{filter} = \delta \sum_{l=1}^{G}\Big(2B(|x_l| + |y_l|) + \frac{2|w_l|}{p} + \nguyen{|bi_l|}\Big)
\end{split}\end{equation}

\subsubsection{Channel parallelism}
Similar to the filter parallelism, channel parallelism splits the DL models horizontally, i.e., by the number of input channels $C$.
Thus, the computation time, and the required memory at each PE are same as those of filter parallelism %(Equations~\ref{equal:channel_comp},~\ref{equal:channel_comm},~\ref{equal:channel_mem}).
\begin{equation}
\scriptsize
\label{equal:channel_mem}
\begin{split}
M_{channel} = \delta \sum_{l=1}^{G}\Big(2B(|x_l| + |y_l|) + \frac{2|w_l|}{p} + \nguyen{|bi_l|} \Big)
\end{split}\end{equation}
\begin{equation}
\label{equal:channel_comp}
\scriptsize
\begin{split}
T_{channel,comp} = T_{filter,comp} = \frac{D}{p}\sum_{l = 1}^{G}\Big(FW_l + BW_l\Big) + \frac{D}{pB}\sum_{l = 1}^{G}(WU_l)
\end{split}
\end{equation}
The communication is performed in a different pattern that includes
(1) an Allreduce at the forward phase (except layer G) that costs $T_{ar}(p,B|y_l|)$,
%$2(p-1)(\alpha + \frac{B|y_l|}{p}\delta\beta)$
%%\begin{equation}
%%\small
%%2(p-1)(\alpha + \frac{B|y_l|}{p}\delta\beta)
%%\end{equation}
and (2) an Allgather at the backward phase (except layer 1) that costs $T_{ag}(p,\frac{B|\frac{dL}{dx_l}|}{p})$.
%$(p-1)(\alpha + \frac{B|\frac{dL}{dx_l}|}{p}\delta\beta)$.
%%\begin{equation}
%%\small
%%(p-1)(\alpha + \frac{B|\frac{dL}{dx_l}|}{p}\delta\beta) = (p-1)(\alpha + \frac{B|x_l|}{p}\delta\beta)
%%\end{equation}
Similar to filter parallelism, we get the total communication time:
\begin{equation}
\scriptsize
\begin{split}
\label{equal:channel_time}
%T_{channel} = \frac{D}{p}\sum_{l=1}^{G}(FW_l + BW_l) 
%+ 3\frac{D}{B}\sum_{l=1}^{G-1} \Big((p-1)(\alpha + \frac{B|y_l|}{p}\delta\beta)\Big)
T_{channel,comm} = 3\frac{D}{B}(p-1)\sum_{l=1}^{G-1} (\alpha + \frac{B|y_l|}{p}\delta\beta)
\end{split}
\end{equation}
\subsubsection{Hybrid parallelism (Data + Filter)}
We consider an example of hybrid parallelism: the combination of data and filter parallelism in which we use $p1$ data parallelism groups in $p = p1 \times p2$ PEs.
We apply filter parallelism inside each group and data parallelism between groups.
Each group will process a partition of the dataset, i.e., $\frac{D}{p1}$ samples.
Each PE then keeps one part of filters of each layer, e.g.,  $\frac{F}{p2}$ filters, so that the required memory is: % summarized in Equation~\ref{equal:hybrid_mem}.
%$2\delta \sum_{l=1}^{G}(\frac{B}{p1}(|x_l| + |y_l|) + \frac{|w_l|}{p2})$. 
%%NguyenTT: Comment for reused later. Uncomment for revision
\begin{equation}
\scriptsize
\label{equal:hybrid_mem_1}
\begin{split}
M_{df} = \delta \sum_{l=1}^{G}\Big(\frac{2B}{p1}(|x_l| + |y_l|) + \frac{2|w_l|}{p2} + \nguyen{|bi_l|}\Big)
\end{split}\end{equation} 

Each PE hence performs $\frac{1}{p2}$ of the computation at each layer with a mini-batch of $\frac{B}{p1}$. The computation time is: % $\sum_{1}^{I} \frac{B}{p1}\sum_{l = 1}^{G}(\frac{FW_l}{p2} + \frac{BW_l}{p2}) 
% $ = \frac{D}{p}\sum_{l = 1}^{G}(FW_l + BW_l)$.
%%NguyenTT: Comment for reused later. Uncomment for revision
\begin{equation}
\label{equal:hybrid_comp1}
\scriptsize
\begin{split}
T_{df,comp} = \sum_{1}^{I} \frac{B}{p1}\sum_{l = 1}^{G}\Big(\frac{FW_l}{p2} + \frac{BW_l}{p2}\Big) + \sum_{1}^{I}\sum_{l=1}^{G}(\frac{WU_l}{p2})
\\
= \frac{D}{p}\sum_{l = 1}^{G}\Big(FW_l + BW_l\Big) + \frac{D}{Bp2}\sum_{l=1}^{G}(WU_l)
\end{split}
\end{equation}

In this strategy, the communication includes intra-group and inter-group communication, which correspond to the cases of filter and data parallelism. The total communication time of one iteration includes $ T_{ag}(p_2,\frac{B|y_l|}{p2})$  and $T_{ar}(p_2,B|\frac{dL}{dx_l}|)$ at each layer and $T_{ar}(p_1,\sum_{l=1}^{G}\frac{|w_l|}{p2}|)$ when update, respectively.
%$3(p2-1)\sum_{l=1}^{G-1}(\alpha + \frac{B}{p1}\frac{|y_l|}{p2}\delta\beta)$ and $2(p1-1)(\alpha + \frac{\sum_{l=1}^{G}\frac{|w_l|}{p2}}{p1}\delta\beta)$, respectively.
% \begin{equation}
%\small
%\begin{split}
%3(p2-1)\sum_{l=1}^{G-1}(\alpha + \frac{B}{p1}\frac{|y_l|}{p2}\delta\beta) + 2(p1-1)(\alpha + \frac{\sum_{l=1}^{G}\frac{|w_l|}{p2}}{p1}\delta\beta)
%\end{split}
%\end{equation}
The total communication time becomes: %is summarized in Equation~\ref{equal:hybrid_comm}.
%%NguyenTT: Comment for reused later
\begin{equation}
\label{equal:hybrid_comm}
\scriptsize
\begin{split}
T_{hybrid,comm} = 3\frac{D}{B}(p2-1)\sum_{l=1}^{G-1}(\alpha + \frac{B|y_l|}{p}\beta) + \\2\frac{D}{B}(p1-1)(\alpha + \frac{\sum_{l=1}^{G}|w_l|}{p}\beta)
%\frac{D}{B}\Big( 3(p2-1)(G-1) + 2(p1-1)\Big)\alpha + \\ \frac{D}{B}\frac{3(p2-1)B\sum_{l=1}^{G-1}|y_l| + 2(p1-1)\sum_{l=1}^{G}|w_l|}{p}\beta) 
\end{split}
\end{equation}
We summarize our analytical model in Table~\ref{table:sum_time}.

\end{document}